\documentclass[aps]{revtex4}

\usepackage{amsmath}
\usepackage{amssymb}
\usepackage{stmaryrd}
\usepackage{natbib}
\numberwithin{equation}{section}
\usepackage{graphicx}

\usepackage{epsfig}

\def\ov#1{\overline{#1}}

\def\wt#1{\widetilde{#1}}
\def\vb#1{\mbox{\boldmath$#1$}}
\def\pd#1#2{\frac{\partial #1}{\partial #2}}

\def\wh#1{\widehat{#1}}
\def\bdot{\,\vb{\cdot}\,}
\def\btimes{\,\vb{\times}\,}

\def\cal#1{\mathcal{#1}}

\newcommand{\bc}{\begin{center}}
\newcommand{\ec}{\end{center}}
\newcommand{\bt}{\begin{tabbing}}
\newcommand{\et}{\end{tabbing}} 
\newcommand{\be}{\begin{eqnarray*}}
\newcommand{\ee}{\end{eqnarray*}}
\newcommand{\bs}{\begin{slide}}
\newcommand{\es}{\end{slide}}

\renewcommand{\theequation}{\arabic{section}.\arabic{equation}}

\begin{document}

\title{Hamiltonian Formulations of Quasilinear Theory for Magnetized Plasmas}

\author{Alain J.~Brizard$^{1,a}$ and Anthony A.~Chan$^{2}$}
\affiliation{$^{1}$Department of Physics, Saint Michael's College, Colchester, Vermont 05439, USA \\ $^{2}$Department of Physics and Astronomy, Rice University, Houston, Texas 77005, USA \\
$^{a}$Author to whom correspondence should be addressed: abrizard@smcvt.edu}

\date{\today}
\def\corrEmail{abrizard@smcvt.edu}

\begin{abstract}
Hamiltonian formulations of quasilinear theory are presented for the cases of uniform and nonuniform magnetized plasmas. First, the standard quasilinear theory of Kennel and Engelmann \cite{K_E} is reviewed and reinterpreted in terms of a general Hamiltonian formulation. Within this Hamiltonian representation, we present the transition from two-dimensional quasilinear diffusion in a spatially uniform magnetized background plasma to three-dimensional quasilinear diffusion in a spatially nonuniform magnetized background plasma based on our previous work \cite{Brizard_Chan:2001,Brizard_Chan:2004}. The resulting quasilinear theory for nonuniform magnetized plasmas yields a $3\times 3$ diffusion tensor that naturally incorporates quasilinear radial diffusion as well as its synergistic connections to diffusion in two-dimensional invariant velocity space (e.g., energy and pitch angle).
\end{abstract}

\maketitle

\tableofcontents

\section{Introduction}

The complex interaction between charged particles and electromagnetic-field wave fluctuations in a magnetized plasma represents a formidable problem with crucial implications toward our understanding of magnetic confinement in laboratory and space plasmas \cite{Kaufman:2019}. These wave-particle interactions can be described either linearly, quasi-linearly, or nonlinearly, depending on how the background plasma is affected by the fluctuating wave fields and the level of plasma turbulence associated with them \cite{Davidson:1972}.

In linear plasma wave theory \cite{Stix:1992}, where the field fluctuations are arbitrarily small, the linearized perturbed Vlasov distribution of each charged-particle species describes the charged-particle response to the presence of small-amplitude electromagnetic waves which, when coupled to the linearized Maxwell wave equations, yields a wave spectrum that is supported by the uniform background magnetized plasma \cite{Stix:1992}. 

In weak plasma turbulence theory \cite{Sagdeev_Galeev:1969,Galeev_Sagdeev:1983}, the background plasma is considered weakly unstable so that a (possibly discrete) spectrum of field perturbations grow to finite but small amplitudes. While these small-amplitude fluctuations interact weakly among themselves, they interact strongly with resonant particles, which satisfy a wave-particle resonance condition in particle phase space (described in terms of unperturbed particle orbits). These resonant wave-particle interactions, in turn, lead to a quasilinear modification of the background Vlasov distribution on a long time scale compared to the fluctuation time scale \cite{Kaufman_JPP:1972,Dewar:1973}. 

Lastly, in strong plasma turbulence theory \cite{Dupree:1966}, nonlinear wave-wave and wave-particle-wave interactions cannot be neglected, and wave-particle resonances include perturbed particle orbits \cite{Galeev_Sagdeev:1983}. The reader is referred to a pedagogical review by Krommes \cite{Krommes:2002} on the theoretical foundations of plasma turbulence as well as a recent study on the validity of quasilinear theory \cite{Crews_Shumlak:2022}. In addition, the mathematical foundations of quasilinear theory for inhomogeneous plasma can be found in the recent work by Dodin \cite{Dodin:2022}.

\subsection{Motivation for this work} 

The primary purpose of the present paper is to present complementary views of two-dimensional quasilinear diffusion in a uniform magnetized plasma. First, we review the quasilinear theory derived by Kennel and Engelmann \cite{K_E}, which represents the paradigm formulation upon which many subsequent quasilinear formulations are derived \cite{Stix:1992}. (We mainly focus our attention on non-relativistic quasilinear theory in the text and summarize the extension to relativistic quasilinear theory in App.~A.) As an alternative formulation of quasilinear theory, we present a Hamiltonian formulation that relies on the use of guiding-center theory for a uniform magnetic field \cite{Cary_Brizard:2009}. In this Hamiltonian formulation, the quasilinear diffusion equation is described in terms of a diffusion tensor whose structure is naturally generalized to three-dimensional quasilinear diffusion in a nonuniform magnetized plasma, as shown in the works of Brizard and Chan \cite{Brizard_Chan:2001,Brizard_Chan:2004}.

Next, two formulations of three-dimensional quasilinear theory are be presented. First, we present a generic quasilinear formulation based on the action-angle formalism \cite{Kaufman_PF:1972,Mahajan_Chen:1985}, which applies to general magnetic-field geometries. This formulation is useful in highlighting the modular features of the quasilinear diffusion tensor. Our second three-dimensional quasilinear formulation is developed for the case of an axisymmetric magnetic field ${\bf B}_{0} = \nabla\psi\btimes\nabla\varphi$, for which the drift action ${\sf J}_{\rm d} = q\psi/c$ is expressed simply in terms of the magnetic flux $\psi$. The presentation of this case is based on a summary of the non-relativistic limit of our previous work \cite{Brizard_Chan:2004}.

\subsection{Notation for quasilinear theory in a uniform magnetized plasma} 

In a homogeneous magnetic field ${\bf B}_{0} = B_{0}\,\wh{\sf z}$, the unperturbed Vlasov distribution $f_{0}({\bf v})$ (for a charged-particle species with charge $q$ and mass $M$) is a function of velocity ${\bf v}$ alone and the perturbed Vlasov-Maxwell fields 
$(\delta f, \delta{\bf E}, \delta{\bf B})$ can be decomposed in terms of Fourier components: $\delta f = \delta\wt{f}({\bf v})\,\exp(i\vartheta) + {\rm c.c.}$ and $(\delta{\bf E}, \delta{\bf B}) = (\delta\wt{\bf E},\delta\wt{\bf B})\, \exp(i\vartheta) + {\rm c.c.}$, where the wave phase is $\vartheta({\bf x},t) = {\bf k}\bdot{\bf x} - \omega\,t$ and the dependence of the eikonal (Fourier) amplitudes $(\delta\wt{f},\delta\wt{\bf E},\delta\wt{\bf B})$ on $({\bf k},\omega)$, which is denoted by a tilde, is hidden. According to Faraday's law, we find $\delta\wt{\bf B} = ({\bf k}c/\omega)\btimes\delta\wt{\bf E}$, which implies ${\bf k}\bdot\delta\wt{\bf B} = 0$. For the time being, however, we will keep the perturbed electric and magnetic fields separate, and assume that the uniform background plasma is perturbed by a monochromatic wave with definite wave vector ${\bf k}$ and wave frequency $\omega$.

Following the notation used by Kennel and Engelmann \cite{K_E}, the velocity ${\bf v}$ and wave vector ${\bf k}$ are decomposed in terms of cylindrical components
\begin{equation}
\left. \begin{array}{rcl}
{\bf v} &=& v_{\|}\,\wh{\sf z} \;+\; v_{\bot}\left(\cos\phi\,\wh{\sf x} + \sin\phi\,\wh{\sf y}\right) \\
{\bf k} &=& k_{\|}\,\wh{\sf z} \;+\; k_{\bot}\left(\cos\psi\,\wh{\sf x} + \sin\psi\,\wh{\sf y}\right)
\end{array} \right\},
\end{equation} 
so that ${\bf k}\bdot{\bf v} = k_{\|}v_{\|} + k_{\bot}v_{\bot}\cos(\phi - \psi)$, where $\phi$ is the gyroangle phase and $\psi$ is the wave-vector phase. We note that the unperturbed Vlasov equation $\partial f_{0}/\partial\phi = 0$ implies that $f_{0}({\bf v})$ is independent of the gyroangle $\phi$, i.e., $f_{0}(v_{\|}, v_{\bot})$. In what follows, we will use the definition
\begin{equation}
   \frac{{\bf k}_{\bot}}{k_{\bot}} \;=\;  \cos\psi\,\wh{\sf x} + \sin\psi\,\wh{\sf y} \;=\; \frac{1}{2}\,e^{i\psi}
    (\wh{\sf x} - i\,\wh{\sf y}) \;+\; \frac{1}{2}\,e^{-i\psi}
    (\wh{\sf x} + i\,\wh{\sf y}) \;\equiv\; \frac{1}{\sqrt{2}} \left(\wh{\sf K} 
    \;+\frac{}{} \wh{\sf K}^{*}\right), 
\label{eq:K_def}
\end{equation}
and the identity
\begin{equation}
 \frac{{\bf v}_{\bot}}{v_{\bot}} \;\equiv\; \wh{\bot} \;=\; \cos\phi\,\wh{\sf x} + \sin\phi\,\wh{\sf y} \;\equiv\; e^{i(\phi - \psi)}\wh{\sf K}/\sqrt{2} \;+\; e^{-i(\phi - \psi)}\wh{\sf K}^{*}/\sqrt{2}.
\label{eq:id}
\end{equation}
We note that, in the work of Kennel and Engelmann \cite{K_E}, the right-handed polarized electric field is $\delta\wt{E}_{R} \equiv \delta\wt{\bf E}\bdot\wh{\sf K}\,e^{-i\psi}$ and the left-handed polarized electric field is $\delta\wt{E}_{L} \equiv \delta\wt{\bf E}\bdot\wh{\sf K}^{*}\,e^{i\psi}$; we will refrain from using these components in the present work.

\section{\label{sec:KE}Kennel-Engelmann Quasilinear Diffusion Equation}

In this Section, we review the quasilinear theory presented by Kennel and Engelmann \cite{K_E} for the case of a uniform magnetized plasma. Here, we make several changes in notation from Kennel and Engelmann's work in preparation for an alternative formulation presented in Sec.~\ref{sec:HQL}.

\subsection{First-order perturbed Vlasov equation}

The linearized perturbed Vlasov equation is expressed in terms of the first-order differential equation for the eikonal amplitude $\delta\wt{f}({\bf v})$:
\begin{equation}
-i \left(\omega - {\bf k}\bdot{\bf v}\right)\delta\wt{f} - \Omega\,\pd{\delta\wt{f}}{\phi} \equiv -\Omega\; e^{i\Theta}\pd{}{\phi}\left( e^{-i\Theta}\;\delta\wt{f}\right) \;=\; -\,\frac{q}{M} \left( \delta\wt{\bf E} + \frac{\bf v}{c}\btimes\delta\wt{\bf B}\right)\bdot
\pd{f_{0}}{\bf v}  
\label{eq:Vlasov_eq}
 \end{equation}
 where $\Omega = qB_{0}/(Mc)$ denotes the (signed) gyrofrequency and the solution of the integrating factor $\partial\Theta/\partial\phi \equiv \Omega^{-1}d\vartheta/dt = ({\bf k}\bdot
 {\bf v} - \omega)/\Omega$ yields
\begin{eqnarray}
\Theta(\phi) &=& \left( \frac{k_{\|}v_{\|} - \omega}{\Omega}\right)\phi \;+\; \frac{k_{\bot}v_{\bot}}{\Omega}\;\sin(\phi - \psi) \;\equiv\; \varphi(\phi) \;+\; \lambda\;\sin(\phi - \psi),
 \label{eq:theta}
\end{eqnarray}
where $\lambda = k_{\bot}v_{\bot}/\Omega$. The perturbed Vlasov equation \eqref{eq:Vlasov_eq}  is easily solved as
\begin{equation}
\delta\wt{f}({\bf v}) \;=\; \frac{q\,e^{i\Theta}}{M\Omega}\int^{\phi} e^{-i\Theta^{\prime}} \left(\delta\wt{\bf E} + \frac{{\bf v}^{\prime}}{c}\btimes\delta\wt{\bf B}\right)\bdot\pd{f_{0}}{{\bf v}^{\prime}}\;d\phi^{\prime},
\label{eq:deltaf_def}
\end{equation}
where a prime denotes a dependence on the integration gyroangle $\phi^{\prime}$. Here, we can write the perturbed evolution operator
\begin{equation}
\frac{q}{M\Omega}\left(\delta\wt{\bf E} + \frac{\bf v}{c}\btimes\delta\wt{\bf B}\right)\bdot\pd{}{\bf v} \;\equiv\; \delta \wt{V}_{\|}\;\pd{}{v_{\|}} \;+\; \delta \wt{V}_{\bot}\;\pd{}{v_{\bot}} \;+\; \delta\wt{\phi}\;\pd{}{\phi},
  \label{eq:delta_v}
\end{equation}
which is expressed in terms of the velocity-space eikonal amplitudes
\begin{eqnarray}
\delta \wt{V}_{\|} &=& \frac{q}{M\Omega} \left( \delta\wt{\bf E} \;+\; \frac{{\bf v}_{\bot}}{c}\btimes\delta\wt{\bf B}\right) \bdot\wh{\sf z}, \label{eq:delta_vpar} \\
\delta \wt{V}_{\bot} &=& \frac{q}{M\Omega} \left( \delta\wt{\bf E} \;+\; \frac{v_{\|}\wh{\sf z}}{c}\btimes\delta\wt{\bf B}\right) \bdot\wh{\bot}, \label{eq:delta_vperp} \\
\delta\wt{\phi} &=& \frac{q}{M\Omega}\left( \delta\wt{\bf E} + \frac{v_{\|}
\wh{\sf z}}{c}\btimes\delta\wt{\bf B}\right)\bdot\frac{\wh{\phi}}{v_{\bot}} - \frac{\delta\wt{B}_{\|}}{B_{0}},
\label{eq:delta_phi}
\end{eqnarray}
where $\wh{\phi} = \partial\wh{\bot}/\partial\phi = \wh{\sf z}\btimes\wh{\bot}$. Whenever direct comparison with the work of Kennel and Engelmann \cite{K_E} is needed, we will use Faraday's law to express $\delta\wt{\bf B} = ({\bf k}c/\omega)\btimes\delta\wt{\bf E}$. With this substitution (see App.~\ref{sec:KE_App} for details), for example, we note that Eqs.~\eqref{eq:delta_v}-\eqref{eq:delta_phi} agree exactly with Eq.~(2.12) of Kennel and Engelmann \cite{K_E}.

We now remark that, since $\partial f_{0}(v_{\|},v_{\bot})/\partial\phi$ vanishes, only the first two terms in Eq.~\eqref{eq:delta_v} are non-vanishing when applied to $f_{0}$. Hence, Eq.~\eqref{eq:deltaf_def} contains the integrals 
\begin{eqnarray} 
e^{i\Theta}\int^{\phi} e^{-i\Theta^{\prime}}d\phi^{\prime}, \label{eq:int_z} \\
e^{i\Theta}\int^{\phi} e^{-i\Theta^{\prime}}\wh{\bot}^{\prime}d\phi^{\prime}. \label{eq:int_xy}
\end{eqnarray}
In order to evaluate these integrals, we use the Bessel-Fourier decomposition $e^{i\Theta} = e^{i\varphi}\;\sum_{\ell = -\infty}^{\infty} J_{\ell}(\lambda)\;e^{i\ell(\phi - \psi)}$,
so that the scalar integral \eqref{eq:int_z} becomes 
\begin{equation}
e^{i\Theta}\int^{\phi} e^{-i\Theta^{\prime}}d\phi^{\prime} = \sum_{m,\ell = -\infty}^{\infty} i\,\Delta_{\ell}\;J_{m}(\lambda)\,J_{\ell}(\lambda)\;e^{i(m-\ell)(\phi - \psi)}, 
\label{eq:phi_z}
\end{equation}
where the resonant denominator is
\begin{equation}
\Delta_{\ell} \;\equiv\; \frac{\Omega}{k_{\|}v_{\|} + \ell\,\Omega - \omega},
\label{eq:Delta_ell}
\end{equation}
while, using the identity \eqref{eq:id}, the vector integral \eqref{eq:int_xy} becomes 
\begin{widetext}
\begin{equation}
e^{i\Theta}\int^{\phi} e^{-i\Theta^{\prime}}\wh{\bot}^{\prime}\;d\phi^{\prime} \;=\; \sum_{m,\ell = -\infty}^{\infty}  i\,\Delta_{\ell}\;J_{m}(\lambda)\; \mathbb{J}_{\bot\ell}(\lambda)\; e^{i(m-\ell)(\phi - \psi)},
\label{eq:phi_xy}
\end{equation}
\end{widetext}
where we introduced the vector-valued Bessel function
\begin{equation}
 \mathbb{J}_{\bot\ell}(\lambda) \;\equiv\; \frac{\wh{\sf K}}{\sqrt{2}}\,J_{\ell + 1}(\lambda) \;+\; \frac{\wh{\sf K}^{*}}{\sqrt{2}}\,J_{\ell - 1}(\lambda),
 \label{eq:J_perp_l}
 \end{equation}
 with the identity
 \begin{equation} 
 {\bf k}\bdot \mathbb{J}_{\bot\ell} \;=\; (J_{\ell +1} + J_{\ell - 1})\;k_{\bot}/2 \;=\; (\ell\Omega/v_{\bot})\;J_{\ell},
 \label{eq:rec_Bessel}
 \end{equation} 
 which follows from a standard recurrence relation for Bessel functions. The perturbed Vlasov distribution \eqref{eq:deltaf_def} is thus expressed as
\begin{equation}
\delta\wt{f} \;=\; \sum_{m,\ell} i\,\Delta_{\ell}\;J_{m}(\lambda)\; e^{i(m-\ell)(\phi - \psi)} \left( \delta\wt{V}_{\|\ell}\,\pd{f_{0}}{v_{\|}} \;+\; \delta\wt{V}_{\bot\ell}\,\pd{f_{0}}{v_{\bot}} \right),
\label{eq:deltaf_ml}
\end{equation}
where the Bessel-Fourier components are
\begin{eqnarray}
\delta\wt{V}_{\|\ell} &=& \frac{q}{M\Omega}\,\delta\wt{E}_{\|}\;J_{\ell}(\lambda) \;-\; v_{\bot}\,\wh{\sf z}\btimes\frac{\delta\wt{\bf B}}{B_{0}}\bdot  \mathbb{J}_{\bot\ell}(\lambda),
\label{eq:deltaV_parell} \\
\delta\wt{V}_{\bot\ell} &=& \frac{q}{M\Omega} \left(\delta\wt{\bf E} + \frac{v_{\|}\wh{\sf z}}{c}\btimes\delta\wt{\bf B}\right)\bdot  \mathbb{J}_{\bot\ell}(\lambda).
\label{eq:deltaV_perpell}
\end{eqnarray}
Once again, Eqs.~\eqref{eq:deltaf_ml}-\eqref{eq:deltaV_perpell} agree exactly with Eq.~(2.19) of Kennel and Engelmann \cite{K_E} when Faraday's law is inserted in Eqs.~\eqref{eq:deltaV_parell}-\eqref{eq:deltaV_perpell}; see App.~\ref{sec:KE_App} for details. Th relativistic version of Eqs.~\eqref{eq:deltaf_ml}-\eqref{eq:deltaV_perpell}, which was first derived by Lerche \cite{Lerche:1968}, is also shown in App.~\ref{sec:KE_App}.

\subsection{Quasilinear diffusion in velocity space}

We are now ready to calculate the expression for the quasilinear diffusion equation for the slow evolution ($\tau = \epsilon^{2}t$) of the background Vlasov distribution
\begin{equation}
\frac{1}{\Omega}\pd{f_{0}}{\tau} \;=\; -\;{\rm Re}\left[\left\langle \frac{q}{M\Omega}\left(\delta\wt{\bf E}^{*} + \frac{\bf v}{c}\btimes\delta\wt{\bf B}^{*}\right)\bdot\pd{\delta\wt{f}}{\bf v} \right\rangle \right] \;=\;  -\,{\rm Re}\left[ \left\langle\left(\delta \wt{V}_{\|}^{*}\;\pd{}{v_{\|}} \;+\; 
\delta \wt{V}_{\bot}^{*}\;\pd{}{v_{\bot}} \;+\; \delta\wt{\phi}^{*}\;\pd{}{\phi}\right)\delta\wt{f}\right\rangle \right],
\label{eq:f0_QL}
\end{equation}
where $\epsilon$ denotes the amplitude of the perturbation fields, $\langle\;\rangle$ 
denotes a gyroangle average, and $(\delta\wt{V}_{\|}^{*},\delta\wt{V}_{\bot}^{*}, \delta\wt{\phi}^{*})$ are the complex conjugates of 
Eqs.~\eqref{eq:delta_vpar}-\eqref{eq:delta_phi}. In addition, the real part appears on the 
right side of Eq.~\eqref{eq:f0_QL} as a result of averaging with respect to the wave phase 
$\vartheta$. We note that Kennel and Engelmann \cite{K_E} ignore the term $\partial 
f_{2}/\partial t$ on the left side of Eq.~\eqref{eq:f0_QL}, which is associated with the 
second-order perturbed Vlasov distribution $f_{2}$ generated by non-resonant particles \cite{Kaufman_JPP:1972,Dewar:1973}. While this term was shown by Kaufman \cite{Kaufman_JPP:1972} to be essential in demonstrating the energy-momentum conservation laws of quasilinear theory, it is also omitted here and the right side of Eq.~\eqref{eq:f0_QL} only contains resonant-particle contributions.

First, since Eqs.~\eqref{eq:delta_vpar}-\eqref{eq:delta_vperp} are independent of $v_{\|}$ and $v_{\bot}$, respectively, we find
\begin{eqnarray} 
\left\langle\delta \wt{V}_{\|}^{*}\;\pd{\delta\wt{f}}{v_{\|}} \;+\; \delta \wt{V}_{\bot}^{*}\;\pd{\delta\wt{f}}{v_{\bot}}\right\rangle &=& \pd{}{v_{\|}} \left\langle\delta \wt{V}_{\|}^{*}\frac{}{}\delta\wt{f}\right\rangle \;+\; \pd{}{v_{\bot}} \left\langle\delta \wt{V}_{\bot}^{*}
\frac{}{}\delta\wt{f}\right\rangle \nonumber \\
 &=& \pd{}{v_{\|}} \left\langle\delta \wt{V}_{\|}^{*}\frac{}{}\delta\wt{f}\right\rangle \;+\; \frac{1}{v_{\bot}}\pd{}{v_{\bot}} \left( v_{\bot}\;\left\langle\delta \wt{V}_{\bot}^{*}\frac{}{}\delta\wt{f}\right\rangle \right) \;-\; \left\langle\frac{\delta \wt{V}_{\bot}^{*}}{v_{\bot}}\;\delta\wt{f}\right\rangle,
 \label{eq:div_id}
 \end{eqnarray}
where we took into account the proper Jacobian $(v_{\bot})$ in cylindrical velocity space $(v_{\|}, v_{\bot}, \phi)$. On the other hand, the third term in Eq.~\eqref{eq:f0_QL} can be written as
\[ \left\langle \delta\wt{\phi}^{*}\;\pd{\delta\wt{f}}{\phi}\right\rangle \;=\; -\; \left\langle \left(\pd{\delta\wt{\phi}^{*}}{\phi}\right)\;\delta\wt{f}\right\rangle \;=\;  \frac{q}{M\Omega}\left( \delta\wt{\bf E}^{*} + \frac{v_{\|}\wh{\sf z}}{c}\btimes\delta\wt{\bf B}^{*}\right)\bdot
\left\langle \frac{\wh{\bot}}{v_{\bot}}\;\delta\wt{f}\right\rangle \;\equiv\; \left\langle\frac{\delta \wt{V}_{\bot}^{*}}{v_{\bot}}\;\delta\wt{f}\right\rangle, \]
where the last term in Eq.~\eqref{eq:delta_phi} is independent of the gyroangle $\phi$. Since this term cancels the last term in Eq.~\eqref{eq:div_id}, the quasilinear diffusion equation \eqref{eq:f0_QL} becomes
\begin{equation}
\frac{1}{\Omega}\,\pd{f_{0}}{\tau} \;=\; -\;\pd{}{v_{\|}} \left({\rm Re}\left\langle \delta \wt{V}_{\|}^{*}\frac{}{}\delta\wt{f}\right\rangle \right) \;-\; \frac{1}{v_{\bot}} \pd{}{v_{\bot}} \left( v_{\bot}\;{\rm Re}\left\langle \delta \wt{V}_{\bot}^{*}\frac{}{}\delta\wt{f}\right\rangle \right).
\label{eq:QL_eq}
\end{equation}

Next, using the identity \eqref{eq:id}, we find 
\[ \sum_{m} J_{m}(\lambda)\left\langle \left(1,\frac{}{} \wh{\bot}\right)
\;e^{i(m-\ell)(\phi - \psi)}\right\rangle \;=\; \left(J_{\ell}(\lambda),\frac{}{} \mathbb{J}_{\bot\ell}(\lambda)\right), \]
so that, from Eq.~\eqref{eq:deltaf_ml}, we find
\begin{eqnarray}
\sum_{m} J_{m}(\lambda)\left\langle\delta \wt{V}_{\|}^{*}\frac{}{}e^{i(m-\ell)(\phi - \psi)}\right\rangle &=& \frac{q}{M\Omega}\,\delta\wt{E}_{\|}^{*}\;J_{\ell}(\lambda)  \;-\; v_{\bot}\,\wh{\sf z}\btimes\frac{\delta\wt{\bf B}^{*}}{B_{0}}\bdot \mathbb{J}_{\bot\ell}(\lambda) \;\equiv\; 
\delta\wt{V}_{\|\ell}^{*}, \\
\sum_{m} J_{m}(\lambda)\left\langle\delta \wt{V}_{\bot}^{*}\frac{}{}e^{i(m-\ell)(\phi - \psi)}\right\rangle &=& \frac{q}{M\Omega} \left(\delta\wt{\bf E}^{*} + \frac{v_{\|}\wh{\sf z}}{c}\btimes\delta\wt{\bf B}^{*}\right)\bdot\mathbb{J}_{\bot\ell}(\lambda) \;\equiv\; 
\delta\wt{V}_{\bot\ell}^{*}.
\end{eqnarray}
Hence,  the quasilinear diffusion equation \eqref{eq:QL_eq} can be written as
\begin{eqnarray}
\frac{1}{\Omega}\pd{f_{0}}{\tau} &=& -\; \pd{}{v_{\|}} \left\{ {\rm Re}\left[\sum_{\ell = -\infty}^{\infty}  i\,\Delta_{\ell}\,\delta\wt{V}_{\|\ell}^{*}  \left( \delta\wt{V}_{\|\ell}\,\pd{f_{0}}{v_{\|}} \;+\; \delta\wt{V}_{\bot\ell}\,\pd{f_{0}}{v_{\bot}} \right)\right] \right\} \nonumber \\
 &&-\; \frac{1}{v_{\bot}}\pd{}{v_{\bot}}  \left\{ v_{\bot}\;{\rm Re}\left[  \sum_{\ell = -\infty}^{\infty}  i\,\Delta_{\ell}\,\delta\wt{V}_{\bot\ell}^{*} \left( \delta\wt{V}_{\|\ell}\,\pd{f_{0}}{v_{\|}} \;+\; \delta\wt{V}_{\bot\ell}\,\pd{f_{0}}{v_{\bot}} \right)\right] \right\} 
 \;\equiv\; \pd{}{\bf v}\bdot\left( \vb{\sf D}\bdot\pd{f_{0}}{\bf v} \right),
  \label{eq:QL_eq_vv}
\end{eqnarray}
where the diagonal diffusion coefficients are
\begin{eqnarray}
{\sf D}^{\|\|} &\equiv& \wh{\sf z}\bdot\vb{\sf D}\bdot\wh{\sf z} \;=\; \sum_{\ell = -\infty}^{\infty} \;{\rm Re}\left(-\, i\,\Delta_{\ell}\right)\; |\delta\wt{V}_{\|\ell}|^{2}, \label{eq:D_parpar} \\
{\sf D}^{\bot\bot} &\equiv& \wh{\bot}\bdot\vb{\sf D}\bdot\wh{\bot} \;=\;  \sum_{\ell = -\infty}^{\infty} \;{\rm Re}\left(-\; i\,\Delta_{\ell}\right)\;|\delta\wt{V}_{\bot\ell}|^{2}, \label{eq:D_perper}
\end{eqnarray}
while the off-diagonal diffusion coefficients are
\begin{eqnarray}
{\sf D}^{\|\bot} &\equiv& \wh{\sf z}\bdot\vb{\sf D}\bdot\wh{\bot} \;=\;  \sum_{\ell = -\infty}^{\infty} {\rm Re}\left(-\; i\,\Delta_{\ell}\right)\;{\rm Re}(
\delta\wt{V}_{\|\ell}^{*}\, \delta\wt{V}_{\bot\ell}), \label{eq:D_parper} \\
{\sf D}^{\bot\|} &\equiv& \wh{\bot}\bdot\vb{\sf D}\bdot\wh{\sf z} \;=\; \sum_{\ell = -\infty}^{\infty} {\rm Re}\left(-\; i\,\Delta_{\ell}\right)\;{\rm Re}(
\delta\wt{V}_{\bot\ell}^{*}\, \delta\wt{V}_{\|\ell}), \label{eq:D_perpar}
\end{eqnarray}
which are defined to be explicitly symmetric (i.e., ${\sf D}^{\|\bot} = {\sf D}^{\bot\|}$). Here, using the Plemelj formula 
\cite{Stix:1992}, we find
\begin{equation}
{\rm Re}\left(-\; i\,\Delta_{\ell}\right) \;=\; {\rm Re}\left[ \frac{i\,\Omega}{(\omega - k_{\|}v_{\|} - \ell\Omega)}\right] \;=\; \pi\,\Omega\;\delta(\omega_{r} \;-\; k_{\|}v_{\|} - \ell\Omega), 
\label{eq:Plemelj}
\end{equation}
where we assumed $\omega = \omega_{r} + i\,\gamma$ and took the weakly unstable limit $\gamma \rightarrow 0^{+}$. Hence, the quasilinear diffusion coefficients \eqref{eq:D_parpar}-\eqref{eq:D_perpar} are driven by resonant particles, which satisfy the resonance condition $k_{\|} v_{\|{\rm res}} \equiv \omega - \ell\Omega$. The reader is referred to the early references by Kaufman \cite{Kaufman_JPP:1972} and Dewar \cite{Dewar:1973} concerning the role of non-resonant particles in demonstrating the energy-momentum conservation laws of quasilinear theory.

Equation (2.25) from Kennel and Engelmann \cite{K_E} (see App.~\ref{sec:KE_App}) can be expressed as the dyadic diffusion tensor
\begin{equation}
\vb{\sf D} \;\equiv\; \sum_{\ell} {\rm Re}(-\, i\,\Delta_{\ell})\; 
\wt{\bf v}_{\ell}^{*}\,\wt{\bf v}_{\ell} \;=\;  \sum_{\ell} {\rm Re}(-\, i\,\Delta_{\ell}) \left[\left(\delta\wt{V}^{*}_{\|\ell}\;\wh{\sf z} \;+\; \delta\wt{V}^{*}_{\bot\ell}\;\wh{\bot}\right)\frac{}{}
\left(\delta\wt{V}_{\|\ell}\;\wh{\sf z} \;+\; \delta\wt{V}_{\bot\ell}\;\wh{\bot}\right)\right],
\label{eq:diffusion_D}
\end{equation}
which is Hermitian since the term $- i\,\Delta_{\ell}$ is replaced with ${\rm Re}(-\, i\,\Delta_{\ell})$. Here, the perturbed velocity
\begin{eqnarray}
\wt{\bf v}_{\ell} = \delta\wt{V}_{\|\ell}\;\wh{\sf z} \;+\; \delta\wt{V}_{\bot\ell}\;\wh{\bot} &=& \frac{q\,\delta\wt{\bf E}}{M\Omega}\bdot\left[J_{\ell}(\lambda)\;\wh{\sf z}\,\wh{\sf z} \;+\; \mathbb{J}_{\bot\ell}(\lambda) \; \wh{\bot}\right] \;+\; \wh{\sf z}
\btimes\frac{\delta\wt{\bf B}}{B_{0}}\bdot\mathbb{J}_{\bot\ell}(\lambda)\;\left( v_{\|}\,\wh{\bot} \;-\frac{}{} v_{\bot}\,\wh{\sf z} \right)
 \end{eqnarray}
 explicitly separates the electric and magnetic contributions to the quasilinear diffusion tensor \eqref{eq:diffusion_D}. In particular, the role of the perturbed perpendicular magnetic field is clearly seen in the process of pitch-angle diffusion because of the presence of the terms $(v_{\|}\,\wh{\bot} - v_{\bot}\, \wh{\sf z})$ associated with it. We also note that the parallel component of the perturbed magnetic field, $\delta\wt{B}_{\|} = \wh{\sf z}\bdot\delta\wt{\bf B}$, does not contribute to quasilinear diffusion in a uniform magnetized plasma. The components of the perturbed electric field, on the other hand, involve the parallel component, $\delta\wt{E}_{\|} = \wh{\sf z}\bdot\delta\wt{\bf E}$, as well as the right and left polarized components, $\delta\wt{E}_{R} = \delta\wt{\bf E}\bdot(\wh{\sf x} - i\,\wh{\sf y})/
 \sqrt{2}$ and $\delta\wt{E}_{L} = \delta\wt{\bf E}\bdot(\wh{\sf x} + i\,\wh{\sf y})/
 \sqrt{2}$, respectively, appearing through the definition \eqref{eq:J_perp_l}.
 
 Lastly, we note that the dyadic form \eqref{eq:diffusion_D} of the quasilinear diffusion tensor in the quasilinear diffusion equation \eqref{eq:QL_eq_vv} can be used to easily verify that the unperturbed entropy ${\cal S}_{0} \equiv -\,\int f_{0}\; \ln f_{0}\; d^{3}v$ satisfies the H Theorem:
 \begin{equation}
\frac{d{\cal S}_{0}}{dt} \;=\; -\,\epsilon^{2}\int \pd{f_{0}}{\tau}\;(\ln f_{0} + 1)\;
d^{3}v \;=\; \epsilon^{2} \sum_{\ell}\int {\rm Re}(-i\Delta_{\ell})\;f_{0}\;\left|
\wt{\bf v}_{\ell}\bdot\pd{\ln f_{0}}{\bf v}\right|^{2}\;d^{3}v \;>\; 0.
\label{eq:S0_H}
 \end{equation}
 Once again, the energy-momentum conservation laws in quasilinear theory will not be discussed here. Instead the interested reader can consult earlier references \cite{Kaufman_JPP:1972,Dewar:1973}, as well as Chapters 16-18 in the standard textbook by Stix \cite{Stix:1992}.

\subsection{Quasilinear diffusion in invariant velocity space}

In preparation for Sec.~\ref{sec:HQL}, we note that a natural choice of velocity-space coordinates, suggested by guiding-center theory, involves replacing the parallel velocity $v_{\|}$ with the parallel momentum $p_{\|} = M\,v_{\|}$ and the perpendicular speed $v_{\bot}$ with the magnetic moment $\mu = Mv_{\bot}^{2}/(2B_{0})$. We note that these two coordinates are independent dynamical invariants of the particle motion in a uniform magnetic field.

With this change of coordinates, the quasilinear diffusion equation \eqref{eq:QL_eq_vv} becomes
\begin{equation}
\frac{1}{\Omega}\pd{f_{0}}{\tau} \;\equiv\; \pd{}{p_{\|}} \left( D^{pp}\; \pd{f_{0}}{p_{\|}} \;+\; D^{p\mu}\;\pd{f_{0}}{\mu} \right) \;+\; \pd{}{\mu} \left( D^{\mu p}\; \pd{f_{0}}{p_{\|}} \;+\; D^{\mu\mu}\;\pd{f_{0}}{\mu} \right),
\label{eq:QL_eq_vmu}
\end{equation}
where the quasilinear diffusion coefficients are
\begin{equation}
\left. \begin{array}{rcl}
D^{pp} &=& M^{2}\,{\sf D}^{\|\|} \;=\; \sum_{\ell} {\rm Re}(-i\,\Delta_{\ell})\;|\delta\wt{P}_{\|\ell}|^{2} \\
D^{p\mu} &=& (M^{2}v_{\bot}/B_{0})\,{\sf D}^{\|\bot} \;=\; \sum_{\ell} 
{\rm Re}(-i\,\Delta_{\ell})\;{\rm Re}\left(\delta\wt{P}_{\|\ell}^{*}\frac{}{}\delta\wt{\mu}_{\ell}\right) \\
D^{\mu\mu} &=& (Mv_{\bot}/B_{0})^{2}\,{\sf D}^{\bot\bot} \;=\; \sum_{\ell} 
{\rm Re}(-i\, \Delta_{\ell})\;|\delta\wt{\mu}_{\ell}|^{2} 
\end{array} \right\},
\label{eq:D_pmu}
\end{equation}
with the eikonal amplitudes
\begin{eqnarray}
\delta\wt{P}_{\|\ell} &=& \frac{q}{\Omega} \left( \delta\wt{E}_{\|}\;J_{\ell} \;+\; \frac{v_{\bot}}{c}\,\mathbb{J}_{\bot\ell}\btimes\delta\wt{\bf B}\bdot\wh{\sf z}\right), \label{eq:P_ell} \\
\delta\wt{\mu}_{\ell} &=& \frac{q}{B_{0}\Omega} \left( \delta\wt{\bf E} \;+\; \frac{v_{\|}\,\wh{\sf z}}{c}\btimes\delta\wt{\bf B}\right)\bdot v_{\bot}\,\mathbb{J}_{\bot\ell}, \label{eq:mu_ell}
\end{eqnarray}
and the symmetry $D^{\mu p} = D^{p\mu}$ follows from the assumption of a Hermitian diffusion tensor. Lastly, as expected, we note that the eikonal amplitude for the perturbed kinetic energy
\begin{equation}
  \delta\wt{\cal E}_{\ell} \;\equiv\; M{\bf v}\bdot\wt{\bf v}_{\ell} \;=\; v_{\|}\,
  \delta\wt{P}_{\|\ell} + \delta\wt{\mu}_{\ell}\,B_{0} \;=\; \frac{q}{\Omega}\;
  \delta\wt{\bf E}\bdot\left( v_{\|}\;J_{\ell} \;+\frac{}{} v_{\bot}\; 
  \mathbb{J}_{\bot\ell}\right), 
\end{equation}
only involves the perturbed electric field. Hence, another useful representation of quasilinear diffusion in invariant velocity $({\cal E}, \mu)$ space is given by the quasilinear diffusion equation
\begin{equation}
\frac{1}{\Omega}\pd{f_{0}}{\tau} \;\equiv\; v_{\|}\;\pd{}{\cal E} \left[ \frac{1}{v_{\|}} \left( D^{{\cal E}{\cal E}}\; \pd{f_{0}}{\cal E} \;+\; D^{{\cal E}\mu}\;\pd{f_{0}}{\mu} \right) \right] \;+\; v_{\|}\;\pd{}{\mu} \left[ \frac{1}{v_{\|}} \left( D^{\mu{\cal E}}\; \pd{f_{0}}{\cal E} \;+\; D^{\mu\mu}\;\pd{f_{0}}{\mu} \right) \right],
\label{eq:QL_eq_Emu}
\end{equation}
where the quasilinear diffusion coefficients are
\begin{equation}
\left. \begin{array}{rcl}
D^{{\cal E}{\cal E}} &=& \sum_{\ell} {\rm Re}(-i\,\Delta_{\ell})
\;|\delta\wt{{\cal E}}_{\ell}|^{2} \\
D^{{\cal E}\mu} &=& \sum_{\ell} {\rm Re}(-i\,\Delta_{\ell})\;
{\rm Re}\left(\delta\wt{{\cal E}}_{\ell}^{*}
\frac{}{}\delta\wt{\mu}_{\ell}\right) \\
D^{\mu\mu} &=& \sum_{\ell} {\rm Re}(-i\,\Delta_{\ell})
\;|\delta\wt{\mu}_{\ell}|^{2} 
\end{array} \right\},
\label{eq:D_Emu}
\end{equation}
and the Jacobian $1/v_{\|}$ is a function of $({\cal E},\mu)$: $|v_{\|}| = \sqrt{(2/M)({\cal E} - \mu\,B_{0})}$, while the sign of $v_{\|}$ is a constant of the motion in a uniform 
magnetic field.

\section{\label{sec:HQL}Hamiltonian Quasilinear Diffusion Equation}

In Sec.~\ref{sec:KE}, we reviewed the standard formulation of quasilinear theory in a uniform magnetized plasma \cite{K_E}. In this Section, we introduce the Hamiltonian formulation of the Vlasov equation from which we will derive the Hamiltonian quasilinear diffusion equation, which will then be compared with the Kennel-Engelmann quasilinear diffusion equation \eqref{eq:QL_eq_vv}. 

In order to proceed with a Hamiltonian formulation, however, we will be required to express the perturbed electric and magnetic fields in terms of perturbed electric and magnetic potentials. We note that, despite the use of these potentials, the gauge invariance of the Hamiltonian quasilinear diffusion equation will be guaranteed in the formulation adopted here.

\subsection{Non-adiabatic decomposition of the perturbed Vlasov distribution}

The Hamiltonian formulation of quasilinear diffusion begins with the representation of the perturbed electric and magnetic fields in terms of the perturbed electric scalar potential $\delta\Phi$ and the perturbed magnetic vector potential $\delta{\bf A}$, where 
$\delta{\bf E} = -\nabla\delta\Phi - c^{-1}\partial\delta{\bf A}/\partial t$ and 
$\delta{\bf B} = \nabla\btimes\delta{\bf A}$. Hence, we find the identity
\begin{equation}
\delta{\bf E} \;+\; \frac{\bf v}{c}\btimes\delta{\bf B} \;=\; -\;\nabla\left( \delta\Phi \;-\; \frac{\bf v}{c}\bdot\delta{\bf A}\right) \;-\; \frac{1}{c}\,\frac{d\delta{\bf A}}{dt} \;\equiv\; -\;\nabla\delta\Psi \;-\; \frac{1}{c}\,\frac{d\delta{\bf A}}{dt},
\label{eq:EB_PhiA}
\end{equation}
where $d/dt$ denotes the unperturbed time derivative. We note that the gauge transformation 
\begin{equation}
(\delta\Phi, \delta{\bf A}, \delta\Psi) \;\rightarrow\; \left(\delta\Phi - \frac{1}{c}\pd{\delta\chi}{t},\; \delta{\bf A} + \nabla\delta\chi,\; \delta\Psi \;-\; \frac{1}{c}\,\frac{d\delta\chi}{dt}\right)
\label{eq:gauge}
\end{equation}
guarantees the gauge invariance of the right side of Eq.~\eqref{eq:EB_PhiA}.

Next, by removing the perturbed magnetic vector potential $\delta{\bf A}$ from the canonical momentum 
\[ {\bf P} \;=\; m{\bf v} + q\,({\bf A}_{0} + \epsilon\,\delta{\bf A})/c \;\rightarrow\; {\bf P}_{0} \;=\; m{\bf v} + q{\bf A}_{0}/c, \]
the noncanonical Poisson bracket (which can also be expressed in divergence form)
\begin{eqnarray}
\{ f,\;g\} &=& \frac{1}{M} \left(\nabla f\bdot\pd{g}{\bf v} \;-\; \pd{f}{\bf v}\bdot\nabla g\right) \;+\; \frac{q{\bf B}_{0}}{M^{2}c}\bdot\pd{f}{\bf v}\btimes\pd{g}{\bf v}
\label{eq:PB} \\
 &=& \pd{}{\bf v}\bdot\left[ \frac{1}{M}\left(\nabla f \;+\; 
\frac{q{\bf B}_{0}}{Mc} \btimes\pd{f}{\bf v}\right)\;g \right] \;-\; \nabla\bdot\left( 
\pd{f}{\bf v}\;\frac{g}{M}\right)
\label{eq:PB_div}
\end{eqnarray}
only contains the unperturbed magnetic field ${\bf B}_{0}$., where $f$ and $g$ are arbitrary functions of $({\bf x},{\bf v})$.

The removal of the perturbed magnetic vector potential  $\delta{\bf A}$ from the noncanonical Poisson bracket \eqref{eq:PB}, however, implies that the perturbed Vlasov distribution can be written as
\begin{equation}
\delta f \;=\; \frac{q}{c}\,\delta{\bf A}\bdot\pd{f_{0}}{\bf v} \;+\; \delta g \;\equiv\; \frac{q}{c}\,\delta{\bf A}\bdot\{{\bf x}, f_{0}\} \;+\; \{ \delta s,\; 
f_{0} \},
\label{eq:deltaf_g}
\end{equation}
where the {\it non-adiabatic} contribution $\delta g$ is said to be generated by the perturbation scalar field $\delta s$ \cite{Brizard:1994,Brizard:2018,Brizard_Chandre:2020}, which satisfies the first-order eikonal equation
\begin{equation}
i\;\left({\bf k}\bdot{\bf v} \;-\frac{}{} \omega\right)\delta\wt{s} \;-\; \Omega\;\pd{\delta\wt{s}}{\phi} \;=\; q\,\left( \delta\wt{\Phi} \;-\; 
\frac{\bf v}{c}\bdot\delta\wt{\bf A} \right) \;\equiv\; q\,\delta\wt{\Psi}.
\label{eq:S_eq}
\end{equation} 
Hence, the eikonal solution for $\delta\wt{s}$ is expressed with the same integrating factor used in Eq.~\eqref{eq:deltaf_def}:
\begin{equation}
\delta\wt{s}({\bf v}) \;=\; -\,\frac{q}{\Omega}\;e^{i\Theta}\int^{\phi} \delta\wt{\Psi}^{\prime}\,e^{-i\Theta^{\prime}}\,d\phi^{\prime} \;=\;  -\frac{q}{\Omega}\sum_{m,\ell} i\,\Delta_{\ell}\;J_{m}(\lambda)\; 
e^{i(m-\ell)(\phi - \psi)}\;\delta\wt{\Psi}_{\ell},
\end{equation}
where the gyroangle Fourier component of the effective perturbed potential is
\begin{equation}
\delta\wt{\Psi}_{\ell} \;\equiv\;  \left(\delta\wt{\Phi} - \frac{v_{\|}}{c}\,\delta\wt{A}_{\|}\right) J_{\ell}(\lambda) \;-\; \frac{v_{\bot}}{c}\,\delta\wt{\bf A}\bdot\mathbb{J}_{\bot\ell}(\lambda),
\label{eq:psi_ell}
\end{equation}
 and the eikonal amplitude of the non-adiabatic perturbed Vlasov distribution is
 \begin{equation}
\delta\wt{g} \;=\; e^{-i\vartheta} \left\{ \delta\wt{s}\,e^{i\vartheta},\frac{}{} f_{0}\right\} \;=\; \frac{1}{M} \left( i{\bf k}\,\delta\wt{s} \;+\; \Omega\,\wh{\sf z}\btimes\pd{\delta\wt{s}}{\bf v}\right)\bdot\pd{f_{0}}{\bf v} \;=\; \frac{i{\bf k}}{M} \bdot\pd{f_{0}}{\bf v}\;\delta\wt{s} \;-\; \frac{\Omega}{B_{0}}\,
\pd{\delta\wt{s}}{\phi}\,\pd{f_{0}}{\mu},
 \end{equation} 
where $\mu \equiv M|{\bf v}_{\bot}|^{2}/(2 B_{0})$ denotes the magnetic moment. We note that, under the gauge transformations \eqref{eq:gauge}, the scalar field $\delta s$ transforms as $\delta s \rightarrow \delta s - (q/c)\,\delta\chi$ \cite{Brizard:1994,Brizard:2018,Brizard_Chandre:2020}, and the expression \eqref{eq:deltaf_g} for the perturbed Vlasov distribution is gauge-invariant. Moreover, under the gauge transformation \eqref{eq:gauge}, the eikonal Fourier amplitude \eqref{eq:psi_ell} transforms as
\begin{equation}
    \delta\wt{\Psi}_{\ell} \;\rightarrow\; \delta\wt{\Psi}_{\ell} \;+\; \frac{i}{c} \left( \omega - k_{\|}v_{\|} - \ell\,\Omega\right)\;J_{\ell}\;\delta\wt{\chi},
\label{eq:gauge_ell}
\end{equation}
which is consistent with Eq.~\eqref{eq:gauge}.
 
Next, since the components of the Poisson bracket \eqref{eq:PB} are constant, the unperturbed time derivative of $\delta f$ yields the linearized perturbed Vlasov equation
\begin{eqnarray}
\frac{d\delta f}{dt} &=& \frac{q}{c}\,\frac{d\delta{\bf A}}{dt}\bdot\{{\bf x}, f_{0}\} \;+\;  \frac{q}{c}\,\delta{\bf A}\bdot\{{\bf v}, f_{0}\} \;+\; \left\{ \frac{d\delta s}{dt},\; f_{0} \right\} \nonumber \\
 &=& \frac{q}{c}\,\frac{d\delta{\bf A}}{dt}\bdot\{{\bf x}, f_{0}\} \;+\;  \frac{q}{c}\,\delta{\bf A}\bdot\{{\bf v}, f_{0}\} \;+\; \left\{ q\,\delta\Psi,\; f_{0} \right\} \;=\; \left(  \frac{q}{c}\,\frac{d\delta{\bf A}}{dt} \;+\; q\,\nabla\delta\Psi\right)\bdot\{{\bf x}, f_{0}\} \nonumber \\
  &\equiv& -\,\frac{q}{M} \left( \delta{\bf E} \;+\; \frac{\bf v}{c}\btimes\delta{\bf B}\right)\bdot\pd{f_{0}}{\bf v},
\end{eqnarray}
which implies that the non-adiabatic decomposition \eqref{eq:deltaf_g} is a valid representation of the perturbed Vlasov distribution.

\subsection{Second-order perturbed Vlasov equation}

In order to derive an alternate formulation of quasilinear theory for uniform magnetized plasmas, we begin with second-order evolution of the background Vlasov distribution
\begin{equation}
\pd{f_{0}}{\tau} \;=\; -\frac{q}{M}\left(\delta{\bf E} + \frac{\bf v}{c}\btimes\delta{\bf B}\right)\bdot\pd{\delta f}{\bf v} \;=\; \left(q\,\nabla\delta\Psi \;+\; \frac{q}{c}\,\frac{d\delta{\bf A}}{dt}\right)\bdot\{{\bf x},\; \delta f\},
\label{eq:HQL_prime}
\end{equation}
where, once again, $\tau = \epsilon^{2}t$ denotes the slow quasilinear diffusion time scale, we have ignored the second-order perturbed Vlasov distribution $f_{2}$, and we have inserted Eqs.~\eqref{eq:EB_PhiA} and \eqref{eq:PB}. The first term on the right side of Eq.~\eqref{eq:HQL_prime} can be written as
\begin{eqnarray}
q\nabla\delta\Psi\bdot\{{\bf x},\; \delta f \} &=& \{ q\,\delta\Psi,\; \delta f \} \;-\; q\,\pd{\delta\Psi}{\bf v}\bdot\{{\bf v},\; \delta f\} \;=\;  \{ q\,\delta\Psi,\; \delta f \} \;+\; \frac{q}{c}\,\delta{\bf A}\bdot\{{\bf v},\; \delta f\} \nonumber \\
 &=& \left\{ q\,\delta\Psi,\; \left(\frac{q}{c}\,\delta{\bf A}\bdot\{{\bf x}, f_{0}\} \;+\; \delta g\right) \right\} \;+\;  \frac{q}{c}\,\delta{\bf A}\bdot\left\{{\bf v},\; \delta f\right\},
\end{eqnarray}
where we have inserted the non-adiabatic decomposition \eqref{eq:deltaf_g}, so that the first term can be written as
\begin{eqnarray}
 \left\{ q\,\delta\Psi,\; \frac{q}{c}\,\delta{\bf A}\bdot\{{\bf x}, f_{0}\} \right\} &=& \frac{q^{2}}{c} \left( \{\delta\Psi,\; \delta{\bf A}\}\bdot\{{\bf x},\; f_{0}\} \;+\; \delta{\bf A}\bdot\left\{ \delta\Psi,\frac{}{} \{{\bf x},\; f_{0}\} \right\} \right) \nonumber \\
  &=& \frac{q^{2}}{Mc^{2}}\,\delta{\bf A}\bdot\nabla\delta{\bf A}\bdot\{{\bf x},\; f_{0}\} \;+\; \frac{q}{c}\delta{\bf A}\bdot\left\{ q\,\delta\Psi,\frac{}{} \{{\bf x},\; f_{0}\} \right\},
 \end{eqnarray}
where we used $\{\delta\Psi, \delta{\bf A}\} = (\delta{\bf A}/Mc)\bdot\nabla
\delta{\bf A}$. The second term on the right side of Eq.~\eqref{eq:HQL_prime}, on the other hand, can be written as
\begin{eqnarray}
\frac{q}{c}\,\frac{d\delta{\bf A}}{dt}\bdot\{{\bf x},\; \delta f\} &=& \frac{q^{2}}{c^{2}}\,\frac{d\delta{\bf A}}{dt}\bdot\left\{ {\bf x},\frac{}{} \{{\bf x}, f_{0}\}\right\}\bdot\delta{\bf A} \;+\; \frac{d}{dt}\left( \frac{q}{c}\delta{\bf A}\bdot\{{\bf x},\; \delta g\} \right) \;-\; \frac{q}{c}\delta{\bf A}\bdot\{{\bf v}, \delta g\} \nonumber \\
 &&+\; \frac{q}{c}\delta{\bf A}\bdot\left\{ {\bf x},\frac{}{} \{ f_{0},\; q\,\delta\Psi\}\right\}.
\label{eq:HQL_1}
\end{eqnarray}
Next, by using the Jacobi identity for the Poisson bracket \eqref{eq:PB}:
\begin{equation}
\left\{ f,\frac{}{} \{g,\; h\} \right\} + \left\{ g,\frac{}{} \{h,\; f\} \right\} + \left\{ h,\frac{}{} \{f,\; g\} \right\} \;=\; 0,
\label{eq:Jacobi}
\end{equation}
which holds for arbitrary functions $(f,g,h)$, we obtain
\begin{equation}
\frac{q^{2}}{c}\,\delta{\bf A}\bdot\left( \left\{ \delta\Psi,\frac{}{} \{{\bf x},\; f_{0}\} \right\} \;+\; \left\{ {\bf x},\frac{}{} \{f_{0},\; \delta\Psi \}\right\} \right) \;=\; \frac{q^{2}}{c}\,\delta{\bf A}\bdot\left\{ f_{0},\frac{}{} \{{\bf x},\; \delta\Psi\} \right\} \;\equiv\; -\;\{ f_{0},\;
\delta H_{2}\},
\end{equation}
where $\delta H_{2} = q^{2}|\delta{\bf A}|^{2}/(2 Mc^{2})$ is the second-order perturbed Hamiltonian. We now look at the first term on the right side of Eq.~\eqref{eq:HQL_1}, which we write as
\begin{eqnarray}
\frac{q^{2}}{c^{2}}\,\frac{d\delta{\bf A}}{dt}\bdot\left\{ {\bf x},\frac{}{} \{{\bf x}, f_{0}\}\right\}\bdot\delta{\bf A}  &=& \frac{d}{dt} \left( \frac{q^{2}}{c^{2}}\,\delta{\bf A}\bdot\left\{ {\bf x},\frac{}{} \{{\bf x}, f_{0}\}\right\}\bdot\delta{\bf A}\right) \;-\; \frac{q^{2}}{c^{2}}\,\delta{\bf A}\bdot\left( \left\{ {\bf v},\frac{}{} \{{\bf x}, f_{0}\} \right\} +  \left\{ {\bf x},\frac{}{} \{{\bf v}, f_{0}\} \right\} \right)\bdot\delta{\bf A} \nonumber \\
 &&-\;  \frac{q^{2}}{c^{2}}\,\delta{\bf A}\bdot\left\{ {\bf x},\frac{}{} \{{\bf x}, f_{0}\}\right\}\bdot\frac{d\delta{\bf A}}{dt}.
\end{eqnarray}
Because of  the symmetry of the tensor $\left\{ {\bf x},\frac{}{} \{{\bf x},\; f_{0}\} \right\}$, the last term on the right side (omitting the minus sign) is equal to the left side, so that we obtain
\[ \frac{q^{2}}{c^{2}}\,\frac{d\delta{\bf A}}{dt}\bdot\left\{ {\bf x},\frac{}{} \{{\bf x}, f_{0}\}\right\}\bdot\delta{\bf A} = \frac{d}{dt} \left( \frac{q^{2}}{2c^{2}}\,
\delta{\bf A}\bdot\left\{ {\bf x},\frac{}{} \{{\bf x}, f_{0}\}\right\}\bdot\delta{\bf A}\right) \;-\; \frac{q^{2}}{c^{2}}\,\delta{\bf A}\bdot\left\{ {\bf v},\frac{}{} \{{\bf x}, f_{0}\} \right\}\bdot\delta{\bf A}, \]
where we used the Jacobi identity \eqref{eq:Jacobi} to find $\{{\bf x},\;\{{\bf v}, f_{0}\}\} = \{{\bf v},\;\{{\bf x}, f_{0}\}\}$, since $\{f_{0},\;\{{\bf x}, {\bf v}\}\} = 0$.

When these equations are combined into Eq.~\eqref{eq:HQL_prime}, we obtain the final Hamiltonian form of the second-order perturbed Vlasov equation
\begin{equation}
\pd{f_{0}}{\tau} \;=\; \left\{ \delta H,\frac{}{} \delta g \right\} \;+\; \left\{\delta H_{2},\frac{}{} f_{0}\right\} \;+\; \frac{d}{dt} \left( \frac{q}{c}\,\delta{\bf A}\bdot\{{\bf x},\; \delta g\} \;+\;  \frac{q^{2}}{2c^{2}}\;\delta{\bf A}\bdot\left\{ {\bf x},\frac{}{} \{{\bf x},\; f_{0}\} \right\}
\bdot\delta{\bf A}\right),
\label{eq:HQL_eq}
\end{equation}
where  $\delta H = q\,\delta\Psi = q\,\delta\Phi - q\,\delta{\bf A}\bdot{\bf v}/c$ and $\delta g = \{ \delta s,\; f_{0}\}$.

\subsection{Hamiltonian quasilinear diffusion equation}

We now perform two separate averages of the second-order perturbed Vlasov equation \eqref{eq:HQL_eq}: we first perform an average with respect to the wave phase $\vartheta$, which will be denoted by an overbar, and, second, we perform an average with respect to the gyroangle $\phi$. We begin by noting that the averaged second-order perturbed Hamiltonian $\delta \ov{H}_{2} = q^{2}|\delta\wt{\bf A}|^{2}/(2 Mc^{2})$ is a constant and, therefore, its contribution in Eq.~\eqref{eq:HQL_eq} vanishes upon eikonal-phase averaging. Likewise, the total time derivative in Eq.~\eqref{eq:HQL_eq} vanishes upon eikonal-phase averaging.

The Hamiltonian quasilinear diffusion equation is, therefore, defined as
\begin{eqnarray}
\pd{f_{0}}{\tau} &\equiv& \frac{1}{2}\;\left\langle\ov{\left\{ \delta H,\frac{}{} \delta g \right\}}\right\rangle \;=\; \frac{1}{2}\ov{\left[\nabla\bdot\left( \frac{q}{Mc}\,\delta{\bf A}\;\langle\delta g\rangle \right)\right]} \;+\; \frac{1}{2}\left\langle\pd{}{\bf v}\bdot\ov{\left[ \left( \nabla\delta H \;+\; \Omega\,\frac{q}{c}\,\delta{\bf A}\btimes\wh{\sf z} \right)\;\frac{\delta g}{M}\right]}\right\rangle \nonumber \\
 &=& \frac{1}{2}\left\langle\pd{}{\bf v}\bdot\ov{\left[ \left( \nabla\delta H \;+\; \Omega\,\frac{q}{c}\,\delta{\bf A}\btimes\wh{\sf z} \right)\;\frac{\delta g}{M} \right]}\right\rangle,
 \label{eq:HQL_ave}
\end{eqnarray}
where we used the divergence form \eqref{eq:PB_div} of the Poisson bracket and the eikonal average of the spatial divergence vanishes. Next, the eikonal average of the first term on the last line of the right side of Eq.~\eqref{eq:HQL_ave} yields
\[ \ov{\left(\nabla\delta H\frac{}{} \delta g\right)} \;=\; i\,{\bf k}\;\left(\delta\wt{H}\;\delta\wt{g}^{*} \;-\frac{}{} \delta\wt{H}^{*}\;\delta\wt{g}\right), \]
so that 
\begin{eqnarray} 
\left\langle\pd{}{\bf v}\bdot\ov{\left( \nabla\delta H\;\frac{\delta g}{M}
\right)}\right\rangle &=& \pd{}{p_{\|}} \left( ik_{\|}\frac{}{}
\left\langle\delta\wt{H}\;\delta\wt{g}^{*} \;-\frac{}{} \delta\wt{H}^{*}\;
\delta\wt{g}\right\rangle\right) \;+\; \frac{1}{B_{0}}\pd{}{\mu} \left[ 
i{\bf k}\bdot\left\langle{\bf v}_{\bot}\left(\delta\wt{H}\;\delta\wt{g}^{*} \;-\frac{}{} \delta\wt{H}^{*}\;\delta\wt{g}\right)\right\rangle\right],
 \label{eq:HQL_H}
 \end{eqnarray}
where $p_{\|} = M\,v_{\|}$ and $\mu = M|{\bf v}_{\bot}|^{2}/2B_{0}$. 
The eikonal average of the second term on the last line of the right side of Eq.~\eqref{eq:HQL_ave}, on the other hand, yields
\begin{equation}
 \frac{\Omega}{2\,v_{\bot}}\pd{}{v_{\bot}}\left[\frac{v_{\bot}}{M} \left( \frac{q}{c}\delta\wt{\bf A}\bdot\left\langle\wh{\phi}\,\delta\wt{g}^{*}\right\rangle \;+\; 
\frac{q}{c}\delta\wt{\bf A}^{*}\bdot\left\langle\wh{\phi}\,\delta\wt{g}\right\rangle\right) \right] \;\equiv\; \frac{\Omega}{B_{0}}\pd{}{\mu}\left[ {\rm Re}\left\langle\left( \frac{q}{c}\delta\wt{\bf A}\bdot\pd{{\bf v}_{\bot}}
{\phi}\right) \delta\wt{g}^{*}\right\rangle \right],
\label{eq:HQL_A}
\end{equation}
so that by combining Eqs.~\eqref{eq:HQL_H}-\eqref{eq:HQL_A} into Eq.~\eqref{eq:HQL_ave}, we find
\begin{eqnarray}
\frac{1}{\Omega}\pd{f_{0}}{\tau} &=&  \pd{}{p_{\|}} \left( \frac{k_{\|}}{\Omega}\;{\rm Re}\left\langle i\,\delta\wt{H}\;\delta\wt{g}^{*}\right\rangle\right) \;+\; \frac{1}{B_{0}}\pd{}{\mu}\left[ {\rm Re}\left\langle\left( \frac{q}{c}\delta\wt{\bf A}\bdot\pd{{\bf v}_{\bot}}{\phi} \;+\; i\,\frac{{\bf k}\bdot
{\bf v}_{\bot}}{\Omega}\;\delta\wt{H}\right) \delta\wt{g}^{*} \right\rangle \right],
 \label{eq:HQL}
\end{eqnarray}
In order to evaluate the gyroangle averages in Eq.~\eqref{eq:HQL}, we need to proceed with a transformation from particle phase space to guiding-center phase space, which is presented in the next Section.

\section{Guiding-center Hamiltonian Quasilinear Diffusion Equation}

In this Section, we use the guiding-center transformation \cite{northrop:1963} in order to simplify the calculations involved in obtaining an explicit expression for the Hamiltonian quasilinear diffusion equation \eqref{eq:HQL} that can compared with the standard quasilinear diffusion equation \eqref{eq:QL_eq_vmu} obtained from Kennel-Engelmann's work \cite{K_E}.

\subsection{Guiding-center transformation}

In a uniform background magnetic field, the transformation from particle phase space to guiding-center phase space is simply given as ${\bf x} = {\bf X} + \vb{\rho}$, where the particle position ${\bf x}$ is expressed as the sum of the guiding-center position ${\bf X}$ and the gyroradius vector $\vb{\rho} \equiv \wh{\sf z}\btimes{\bf v}_{\bot}/\Omega$, while the velocity-space coordinates $(p_{\|},\mu,\phi)$ remain unchanged \cite{Cary_Brizard:2009}. Hence, the eikonal wave phase $\vartheta = {\bf k}
\bdot{\bf x} - \omega\,t$ becomes 
\begin{equation}
\vartheta \;=\; {\bf k}\bdot({\bf X} + \vb{\rho}) - \omega\,t \;=\; \theta \;+\; 
{\bf k}\bdot\vb{\rho} \;\equiv\; \theta \;+\; \Lambda,
\label{eq:theta_gc}
\end{equation}
where $\theta$ denotes the guiding-center eikonal wave phase and $\Lambda \equiv \lambda\;\sin(\phi - \psi)$. Next, the particle Poisson bracket \eqref{eq:PB} is transformed into the guiding-center Poisson bracket \cite{Cary_Brizard:2009}
\begin{equation}
\{ F,\; G\}_{\rm gc} \;=\; \wh{\sf z}\bdot\left( \nabla F\;\pd{G}{p_{\|}} \;-\; \pd{F}{p_{\|}}\;\nabla G \right) \;-\; \frac{\Omega}{B_{0}}\left( \pd{F}{\phi}\;\pd{G}{\mu} \;-\; \pd{F}{\mu}\;\pd{G}{\phi} \right) \;-\; 
\frac{c\,\wh{\sf z}}{qB_{0}}\bdot\nabla F\btimes\nabla G,
\label{eq:PB_gc}
\end{equation}
where the last term vanishes in the case of a uniform background plasma since the guiding-center functions $F$ and $G$ depend on the guiding-center position only through the guiding-center wave phase $\theta$ (with $\nabla\theta = {\bf k}$).

\subsection{First-order perturbed guiding-center Vlasov equation}

The guiding-center transformation induces a transformation on particle phase-space functions $f$ to a guiding-center phase-space function $F$ through the guiding-center push-forward ${\sf T}_{\rm gc}^{-1}: F \equiv {\sf T}_{\rm gc}^{-1}f$. For a perturbed particle phase-space function $\delta g \equiv \delta\wt{g}\,\exp(i\vartheta) + {\rm c.c.}$, we find the perturbed guiding-center phase-space function $\delta G \equiv \delta\wt{G}\,\exp(i\theta) + {\rm c.c.}$, where the eikonal amplitude $\delta\wt{G}$ is given by the push-forward expression as
\begin{equation}
\delta\wt{G} \;=\; \delta\wt{g}\; e^{-i\Lambda} \;=\; e^{-i\theta}\;\left\{ \delta\wt{S}\,e^{i\theta},\frac{}{} f_{0} \right\}_{\rm gc} \;=\; i\,k_{\|}\,\pd{f_{0}}{p_{\|}}\;\delta\wt{S} \;-\; \frac{\Omega}{B_{0}}\,\pd{f_{0}}{\mu}\;\pd{\delta\wt{S}}{\phi}.
\label{eq:G_gc}
\end{equation}
The eikonal amplitude of the guiding-center generating function $\delta\wt{S} = \delta\wt{s}\,\exp(-i\Lambda)$ satisfies an equation derived from the first-order eikonal equation \eqref{eq:S_eq}:
\begin{equation}
i\,(k_{\|}v_{\|} - \omega)\;\delta\wt{S} \;-\; \Omega\,\pd{\delta\wt{S}}{\phi} \;=\; \delta\wt{H}\;e^{-i\Lambda} \;\equiv\; \delta\wt{H}_{\rm gc}.
\label{eq:S_gc_eq}
\end{equation}
The solution of the first-order guiding-center eikonal equation \eqref{eq:S_gc_eq} makes use of the gyroangle expansion $\delta\wt{S} = \sum_{\ell = -\infty}^{\infty} \delta\wt{S}_{\ell}\exp[-i\ell(\phi - \psi)]$, which yields the Fourier component
\begin{equation}
\delta\wt{S}_{\ell} \;=\; -\,\frac{i\,\Delta_{\ell}}{\Omega}\;\left\langle \delta\wt{H}\frac{}{}e^{-i\Lambda + i\ell(\phi - \psi)}\right\rangle \;=\; -\,\frac{i\,\Delta_{\ell}}{\Omega}\;q\,\delta\wt{\Psi}_{\ell}.
\label{eq:S_ell}
\end{equation}
Inserting this solution into Eq.~\eqref{eq:G_gc}, with the gyroangle expansion $\delta\wt{G} = \sum_{\ell = -\infty}^{\infty} \delta\wt{G}_{\ell}\exp[-i\ell(\phi - \psi)]$, yields
\begin{equation}
\delta\wt{G}_{\ell} \;=\; i\,\left( k_{\|}\,\pd{f_{0}}{p_{\|}} \;+\; \frac{\ell\,\Omega}{B_{0}}\,\pd{f_{0}}{\mu}\right) \delta\wt{S}_{\ell} \;=\; \frac{q}{\Omega}\,\delta\wt{\Psi}_{\ell}\;\Delta_{\ell}\;\left( k_{\|}\,\pd{f_{0}}{p_{\|}} \;+\; \frac{\ell\,\Omega}{B_{0}}\,\pd{f_{0}}{\mu}\right).
\label{eq:G_ell}
\end{equation}
Hence, the solution for the eikonal amplitude $\delta\wt{g}$ appearing in Eq.~\eqref{eq:HQL} can be obtained from the guiding-center pull-back expression $\delta\wt{g} = \delta\wt{G}\,\exp(i\,\Lambda)$.

\subsection{Guiding-center Hamiltonian quasilinear diffusion equation}

Using the solution \eqref{eq:G_ell} for $\delta\wt{G}_{\ell}$, we are now ready to calculate the quasilinear diffusion equation \eqref{eq:HQL} and obtaina simple dyadic form for the quasilinear diffusion tensor.

\subsubsection{Quasilinear diffusion in guiding-center $(p_{\|},\mu)$-space} 

Now that the solution for the eikonal amplitude $\delta g$ is obtained in terms of the guiding-center phase-space function $\delta\wt{g} = \delta\wt{G}\,\exp(i\,\Lambda)$, we are now able to evaluate the gyroangle-averaged expressions in Eq.~\eqref{eq:HQL}.
We begin with the gyroangle-averaged quadratic term
\begin{eqnarray}
\left\langle\delta\wt{H}\;\delta\wt{g}^{*}\right\rangle &=& \left\langle \delta\wt{H}\; \left(\delta\wt{G}\frac{}{} e^{i\Lambda}\right)^{*}\right\rangle \;=\; \left\langle \left(\delta\wt{H}\frac{}{} e^{-i\Lambda}\right)\;\delta\wt{G}^{*} \right\rangle \;=\; \sum_{\ell = -\infty}^{\infty} \delta\wt{G}_{\ell}^{*} \left\langle \delta\wt{H}\frac{}{} e^{-i\Lambda + i\,\ell(\phi - \psi)}\right\rangle \nonumber \\
 &=& \sum_{\ell = -\infty}^{\infty} q\,\delta\wt{\Psi}_{\ell}\;\delta\wt{G}_{\ell}^{*} \;=\;  \sum_{\ell = -\infty}^{\infty}  \frac{q^{2}}{\Omega}\,|\delta\wt{\Psi}_{\ell}|^{2}\;\Delta_{\ell}^{*}  \left( k_{\|}\;\pd{f_{0}}{p_{\|}} \;+\; \frac{\ell\,\Omega}{B_{0}}\,\pd{f_{0}}{\mu} \right),
\end{eqnarray}
so that
\begin{equation}
\frac{k_{\|}}{\Omega}\;{\rm Re}\left\langle i\,\delta\wt{H}\frac{}{}\delta\wt{g}^{*}\right\rangle \;=\; \sum_{\ell = -\infty}^{\infty} k_{\|}\; {\cal D}_{\ell}  \left( k_{\|}\;\pd{f_{0}}{p_{\|}} \;+\;  \frac{\ell\,\Omega}{B_{0}}\,\pd{f_{0}}{\mu}\right),
\label{eq:p_gcQL}
\end{equation}
where we introduced the quasilinear perturbation potential
\begin{equation}
{\cal D}_{\ell} \;=\; {\rm Re}\left( -\,i\,\Delta_{\ell}\right)\;|(q/\Omega)\,\delta\wt{\Psi}_{\ell}|^{2} \;\equiv\; {\rm Re}\left( -\,i\,\Delta_{\ell}\right)\;
|\delta\wt{\cal J}_{\ell}|^{2},
\label{eq:D_def}
\end{equation}
and
\begin{equation}
 \pd{}{p_{\|}} \left( \frac{k_{\|}}{\Omega}\;{\rm Re}\left\langle i\,\delta\wt{H}\frac{}{}\delta\wt{g}^{*}\right\rangle\right) \;\equiv\; \pd{}{p_{\|}} \left( D_{\rm H}^{pp}\;\pd{f_{0}}{p_{\|}} \;+\; D_{\rm H}^{p\mu}\;\pd{f_{0}}{\mu}\right),
 \label{eq:D_p}
 \end{equation}
 where $D_{\rm H}^{pp} = \sum_{\ell}\,k_{\|}^{2}\,{\cal D}_{\ell}$ and 
 $D_{\rm H}^{p\mu} = \sum_{\ell} k_{\|}\,(\ell\,\Omega/B_{0})\,{\cal D}_{\ell}$.
 
 Next, we find
 \begin{eqnarray}
  \left\langle\left( \frac{q}{c}\delta\wt{\bf A}\bdot\pd{{\bf v}_{\bot}}{\phi} \;+\; i\,\frac{{\bf k}\bdot{\bf v}_{\bot}}{\Omega}\;\delta\wt{H}\right) \delta\wt{g}^{*}\right\rangle &=&  \left\langle\left( \frac{q}{c}\delta\wt{\bf A}
  \bdot\pd{{\bf v}_{\bot}}{\phi} \;+\; i\,
  \frac{{\bf k}\bdot{\bf v}_{\bot}}{\Omega}\;\delta\wt{H}\right)  \left(\delta\wt{G}\frac{}{} e^{i\Lambda}\right)^{*}\right\rangle \nonumber \\
   &=&  \left\langle\left[ \frac{q}{c}\delta\wt{\bf A}\bdot\pd{{\bf v}_{\bot}}{\phi}\,e^{-i\Lambda} \;-\; i\,q\pd{\Lambda}{\phi}\;\left( \delta\wt{\Phi} - \frac{v_{\|}}{c}\delta\wt{A}_{\|} - \frac{{\bf v}_{\bot}}{c}\bdot\delta\wt{\bf A}\right)\,e^{-i\Lambda}\right]
   \delta\wt{G}^{*}\right\rangle \nonumber \\
   &=& -\;\left\langle \pd{}{\phi}\left( \delta\wt{H}\frac{}{} e^{-i\Lambda}\right) \delta\wt{G}^{*}\right\rangle \;=\; \left\langle \delta\wt{H}\;e^{-i\Lambda}\;\pd{\delta\wt{G}^{*}}{\phi}\right\rangle \nonumber \\
    &=& \sum_{\ell = -\infty}^{\infty} i\ell\;\delta\wt{G}_{\ell}^{*} \left\langle \delta\wt{H}\frac{}{} e^{-i\Lambda + i\,\ell(\phi - \psi)}\right\rangle \;=\; \sum_{\ell = -\infty}^{\infty}  i\ell\;q\,\delta\wt{\Psi}_{\ell}\;\delta\wt{G}_{\ell}^{*} \nonumber \\
     &=&  \sum_{\ell = -\infty}^{\infty} i\,\Delta_{\ell}^{*}\;\ell\Omega\;(q/\Omega)^{2}\,|\delta\wt{\Psi}_{\ell}|^{2}\;\left( k_{\|}\,\pd{f_{0}}{p_{\|}} \;+\; \frac{\ell\,\Omega}{B_{0}}\,\pd{f_{0}}{\mu}\right),
 \end{eqnarray}
 so that
 \begin{eqnarray}
\frac{1}{B_{0}}\pd{}{\mu}\left[ {\rm Re}\left\langle\left( \frac{q}{c}\delta\wt{\bf A}\bdot\pd{{\bf v}_{\bot}}{\phi} \;+\; i\, \frac{{\bf k}\bdot{\bf v}_{\bot}}{\Omega}\;\delta\wt{H}\right) \delta\wt{g}^{*} \right\rangle \right] &=& \pd{}{\mu} \left[ \sum_{\ell = -\infty}^{\infty} \frac{\ell\,\Omega}{B_{0}}\;{\cal D}_{\ell}\;\left( k_{\|}\,\pd{f_{0}}{p_{\|}} \;+\; \frac{\ell\,\Omega}{B_{0}}\,\pd{f_{0}}{\mu}\right) \right] \nonumber \\
 &\equiv& \pd{}{\mu} \left( D_{\rm H}^{\mu p}\;\pd{f_{0}}{p_{\|}} \;+\; D_{\rm H}^{\mu\mu}\;\pd{f_{0}}{\mu} \right),
\end{eqnarray}
 where $D_{\rm H}^{\mu p} = \sum_{\ell}\,(\ell\,\Omega/B_{0})\,k_{\|}\,{\cal D}_{\ell}$ and 
 $D_{\rm H}^{\mu\mu} = \sum_{\ell} (\ell\,\Omega/B_{0})^{2}\,
 {\cal D}_{\ell}$.
We can now write the Hamiltonian quasilinear diffusion equation \eqref{eq:HQL} as
 \begin{equation}
\frac{1}{\Omega} \pd{f_{0}}{\tau} \;=\;  \pd{}{p_{\|}} \left( D_{\rm H}^{pp}\;\pd{f_{0}}{p_{\|}} \;+\; D_{\rm H}^{p\mu}\;\pd{f_{0}}{\mu}\right) \;+\; \pd{}{\mu} \left( D_{\rm H}^{\mu p}\;\pd{f_{0}}{p_{\|}} \;+\; D_{\rm H}^{\mu\mu}\;\pd{f_{0}}{\mu} \right).
 \label{eq:HQL_D}
 \end{equation}
 This quasilinear diffusion equation will later be compared with the standard quasilinear diffusion equation \eqref{eq:QL_eq_vmu} derived by Kennel and Engelmann \cite{K_E}.

\subsubsection{Quasilinear diffusion in guiding-center $({\sf J}_{\rm g},{\cal E})$-space} 

Before proceeding with this comparison, however, we consider an alternate representation for the Hamiltonian quasilinear diffusion equation \eqref{eq:HQL_D}, which will be useful in the derivation of a quasilinear diffusion equation for nonuniform magnetized plasmas. If we replace the guiding-center parallel momentum $p_{\|}$ with the guiding-center kinetic energy ${\cal E} = p_{\|}^{2}/2m + \mu\,B_{0}$, and the guiding-center magnetic moment $\mu$ with the gyroaction ${\sf J}_{\rm g} = \mu B_{0}/\Omega$, the Fourier eikonal solution \eqref{eq:G_ell} becomes
\begin{equation}
\delta\wt{G}_{\ell} \;=\; q\,\delta\wt{\Psi}_{\ell}\;
\pd{f_{0}}{\cal E} \;+\; \frac{q}{\Omega}\,
\delta\wt{\Psi}_{\ell}\;\Delta_{\ell} \left( \omega\;\pd{f_{0}}{\cal E} \;+\; \ell\;\pd{f_{0}}{{\sf J}_{\rm g}}\right),
\end{equation}
where the first term on the right side is interpreted as a guiding-center adiabatic contribution to the perturbed Vlasov distribution \cite{Brizard:1994}, while the remaining terms (proportional to the resonant denominator $\Delta_{\ell}$) are non-adiabatic contributions.

By substituting this new solution in Eq.~\eqref{eq:p_gcQL}, we find
\begin{equation}
\frac{k_{\|}}{\Omega}\;{\rm Re}\left\langle i\,\delta\wt{H}\frac{}{}\delta\wt{g}^{*}\right\rangle \;=\; \sum_{\ell = -\infty}^{\infty} k_{\|}\; {\cal D}_{\ell}  \left( \omega\;
\pd{f_{0}}{\cal E} \;+\; \ell\;\pd{f_{0}}{{\sf J}_{\rm g}}\right),
\label{eq:E_gcQL}
\end{equation}
while
\begin{equation}
    {\rm Re}\left\langle\left( \frac{q}{c}\delta\wt{\bf A}\bdot
    \pd{{\bf v}_{\bot}}{\phi} \;+\; i\, \frac{{\bf k}\bdot
    {\bf v}_{\bot}}{\Omega}\;\delta\wt{H}\right) \delta\wt{g}^{*} \right\rangle \;=\; \sum_{\ell = -\infty}^{\infty} \ell\Omega\; 
    {\cal D}_{\ell}  \left( \omega\;\pd{f_{0}}{\cal E} \;+\; \ell\;
    \pd{f_{0}}{{\sf J}_{\rm g}}\right),
\label{eq:mu_gcQL}
\end{equation}
where the guiding-center adiabatic contribution has cancelled out. The guiding-center quasilinear diffusion equation \eqref{eq:HQL_D} becomes
 \begin{equation}
\frac{1}{\Omega} \pd{f_{0}}{\tau} \;=\;  v_{\|}\pd{}{\cal E} \left[ 
\frac{1}{v_{\|}} \left( D_{\rm H}^{{\cal E}{\cal E}}\;\pd{f_{0}}{\cal E} \;+\; 
D_{\rm H}^{{\cal E}{\sf J}}\;\pd{f_{0}}{{\sf J}_{\rm g}}\right)\right] \;+\; v_{\|} 
\pd{}{{\sf J}_{\rm g}} \left[ \frac{1}{v_{\|}} \left( 
D_{\rm H}^{{\sf J}{\cal E}}\;\pd{f_{0}}{\cal E} \;+\; 
D_{\rm H}^{{\sf J}{\sf J}}\;\pd{f_{0}}{{\sf J}_{\rm g}} \right)\right],
 \label{eq:gcHQL_D}
 \end{equation}
 where the guiding-center quasilinear diffusion tensor is represented in $2\times 2$ matrix form as
 \begin{equation}
 \vb{\sf D}_{\rm H} \;\equiv\; \sum_{\ell = -\infty}^{\infty} 
 \left( \begin{array}{cc}
 \ell^{2} & \ell\omega \\
  & \\
 \omega\,\ell & \omega^{2}
 \end{array} \right) {\cal D}_{\ell}.
 \label{eq:D_HQL}
 \end{equation}
 We note that, because of the simple dyadic form of Eq.~\eqref{eq:D_HQL}, other representations for the guiding-center quasilinear diffusion tensor
 $\vb{\sf D}_{\rm H}$ can be easily obtained, e.g., by replacing the guiding-center gyroaction ${\sf J}_{\rm g}$ with the pitch-angle coordinate $\xi = \sqrt{1 - \mu B_{0}/{\cal E}}$. We also note that the dyadic quasilinear tensor \eqref{eq:D_HQL} has a simple modular form compared to the dyadic form \eqref{eq:diffusion_D}.

\subsection{Comparison with Kennel-Engelmann quasilinear theory}

We can now compare the Kennel-Engelmann quasilinear diffusion equation \eqref{eq:QL_eq_vmu} with the guiding-center Hamiltonian quasilinear diffusion equation \eqref{eq:HQL_D}. First, we express the perturbed fields \eqref{eq:P_ell}-\eqref{eq:mu_ell} in terms of the perturbed potentials $(\delta\Phi,\delta{\bf A})$:
\begin{eqnarray}
\delta\wt{P}_{\|\ell} \;=\; M\,\delta\wt{V}_{\|\ell} &=& \frac{q}{\Omega} \left[ J_{\ell}\;\left( -i\,k_{\|}\;\delta\wt{\Phi} \;+\; i\,\frac{\omega}{c}\;\delta\wt{A}_{\|} \right) \;-\; \frac{v_{\bot}}{c} \left( i{\bf k}\;\delta\wt{A}_{\|} \;-\; i\,k_{\|}\;\delta\wt{\bf A}\right)\bdot\mathbb{J}_{\bot\ell} \right] \nonumber \\
 &=& -i\,k_{\|}\;\delta\wt{\cal J}_{\ell} \;+\; i \left(\omega \;-\; k_{\|}v_{\|} \;-\frac{}{} \ell\,\Omega\right) \frac{q\delta\wt{A}_{\|}}{\Omega\,c}\;J_{\ell}, 
\end{eqnarray}
and
\begin{eqnarray}
\delta\wt{\mu}_{\ell} \;=\; \frac{Mv_{\bot}}{B_{0}}\;\delta\wt{V}_{\bot\ell} &=& \frac{q\,v_{\bot}\,\mathbb{J}_{\bot\ell}}{B_{0}\Omega}\bdot \left[ -i\,
{\bf k}\;\delta\wt{\Phi} \;+\; i\,\frac{\omega}{c}\;\delta\wt{\bf A} \;+\; \frac{v_{\|}}{c} \left( i{\bf k}\;\delta\wt{A}_{\|} \;-\; i\,k_{\|}\;\delta\wt{\bf A}\right) \right] 
\nonumber \\
 &=& -i\,\frac{\ell\,\Omega}{B_{0}}\;\delta\wt{\cal J}_{\ell} \;+\; i \left(\omega \;-\; k_{\|}v_{\|} \;-\frac{}{} \ell\,\Omega\right) 
 \frac{q\delta\wt{\bf A}}{cB_{0}\Omega}\bdot v_{\bot}\mathbb{J}_{\bot\ell},
\end{eqnarray}
which are both gauge invariant according to the transformation \eqref{eq:gauge_ell}. Hence, these perturbed fields are expressed in terms of a contribution from the perturbed action $\delta\wt{\cal J}_{\ell}$ and a contribution that vanishes for resonant particles (i.e., $k_{\|}\,v_{\|{\rm res}} = \omega - \ell\Omega$). We note that, in the resonant-particle limit ($\Delta_{\ell} \rightarrow \infty$), the difference between the Kennel-Engelmann formulation and the Hamiltonian formulation vanishes. For example, the Kennel-Engelmann quasilinear diffusion coefficient $D^{pp} = \sum_{\ell} {\rm Re}(-i\Delta_{\ell})\,|\delta\wt{P}_{\|\ell}|^{2}$ is expressed as
\begin{equation}
    D^{pp} \;=\; \sum_{\ell} {\rm Re}(-i\Delta_{\ell}) \left[ k_{\|}^{2}\,
    |\delta\wt{\cal J}_{\ell}|^{2} \;+\; 2 k_{\|}J_{\ell} 
    {\rm Re}\left( \frac{\delta\wt{\cal J}_{\ell}^{*}}{\Delta_{\ell}}\;
    \frac{q\delta\wt{A}_{\|}}{\Omega c}\right) \;+\; \left(\frac{q}{\Omega c}
    \right)^{2} \frac{|\delta\wt{A}_{\|}|^{2}J_{\ell}^{2}}{|\Delta_{\ell}|^{2}} 
    \right] \;\rightarrow\; D_{\rm H}^{pp},
\end{equation}
which yields $D_{\rm H}^{pp}$ in the resonant-particle limit ($\Delta_{\ell} \rightarrow \infty$).

In summary, we have shown that, in the resonant-particle limit ($\Delta_{\ell} \rightarrow \infty$), the Hamiltonian quasilinear diffusion equation \eqref{eq:HQL_D} is identical to the standard quasilinear diffusion equation \eqref{eq:QL_eq_vmu} derived by Kennel and Engelmann \cite{K_E} for the case of a uniform magnetized plasma. In the next Section, we will see how the Hamiltonian quasilinear formalism can be extended to the case of a nonuniform magnetized plasma.

\section{Hamiltonian Quasilinear Formulations for Nonuniform Magnetized Plasma}

In this Section, we briefly review the Hamiltonian formulation for quasilinear diffusion in a nonuniform magnetized background plasma. In an axisymmetric magnetic-field geometry, the $2\times 2$ quasilinear diffusion tensor in velocity space is generalized to a $3\times 3$ quasilinear diffusion tensor that includes radial quasilinear diffusion. In a spatially magnetically-confined plasma, the process of radial diffusion is a crucial element in determining whether charged particles leave the plasma. A prime example is provided by the case of radial diffusion in Earth's radiation belt, which was recently reviewed by Lejosne and Kollmann \cite{Lejosne_Kollmann:2020}.

We present two non-relativistic Hamiltonian formulations of quasilinear diffusion in a nonuniform magnetized plasmas. The first one based on the canonical action-angle formalism \cite{Kaufman_PF:1972,Mahajan_Chen:1985,Mynick_Duvall:1989,Schulz:1996} and the second one based on a summary of our previous work \cite{Brizard_Chan:2004}.

\subsection{\label{subsec:action}Canonical action-angle formalism} 

When a plasma is confined by a nonuniform magnetic field, the charged-particle orbits can be described in terms of 3 orbital angle coordinates $\vb{\theta}$ (generically referred to as the gyration, bounce, and precession-drift angles) and their canonically-conjugate 3 action coordinates ${\bf J}$ (generically referred to as the gyromotion, bounce-motion, and drift-motion actions). In principle, these action coordinates are adiabatic invariants of the particle motion and they are calculated according to standard methods of guiding-center theory \cite{Tao_Chan_Brizard:2007,Cary_Brizard:2009}, which are expressed in terms of asymptotic expansions in powers of a small dimensionless parameter $\epsilon_{B} = \rho/L_{B} \ll 1$ defined as the ratio of a characteristic gyroradius (for a given particle species) and the gradient length scale $L_{B}$ associated with the background magnetic field ${\bf B}_{0}$. When an asymptotic expansion for an adiabatic invariant  ${\sf J} = {\sf J}_{0} + \epsilon_{B}{\sf J}_{1}$ is truncated at first order, for example, we find $d{\sf J}/dt \sim \epsilon_{B}^2$ and the orbital angular average 
$\langle d{\sf J}/dt\rangle = 0$ is the necessary condition for the adiabatic invariance of ${\sf J}$. The reader is referred to Refs.~\cite{Cary_Brizard:2009} and \cite{Tao_Chan_Brizard:2007} where explicit expansions for all three guiding-center adiabatic invariants are derived in the non-relativistic and relativistic limits, respectively, for arbitrary background magnetic geometry.

The canonical action-angle formulation of quasilinear theory assumes that, in the absence of wave-field perturbations, the action coordinates ${\bf J}$ are constants of motion 
$d{\bf J}/dt = -\,\partial H_{0}/\partial\vb{\theta} = 0$, which follows from the 
invariance of the unperturbed Hamiltonian $H_{0}({\bf J})$ on the canonical orbital angles 
$\vb{\theta}$. In this case, the unperturbed Vlasov distribution $F_{0}({\bf J})$ is a function of action coordinates only. We note that the action coordinates considered here are either exact invariants or adiabatic invariants \cite{Kaufman_PF:1972,Mynick_Duvall:1989} of the particle motion, and it is implicitly assumed that any adiabatic action invariant used in this canonical action-angle formulation of quasilinear theory can be calculated to sufficiently high order in $\epsilon_{B}$ within a region of particle phase space that excludes non-adiabatic diffusion in action space \cite{Bernstein:1976}. For example, see Ref.~\cite{Brizard_Markowski:2022} for a brief discussion of the breakdown of the adiabatic invariance of the magnetic moment (on the bounce time scale) for charged particles trapped by an axisymmetric dipole magnetic field. 

In the presence of wave-field perturbations, the perturbed Hamiltonian can be represented in terms of a Fourier decomposition in terms of a discrete wave spectrum $\omega_{k}$ and orbital angles (with Fourier-index vector ${\bf m}$): 
\begin{equation}
    \delta{\cal H}({\bf J},\vb{\theta},t) \;=\; \sum_{{\bf m},k}\;
    \delta\wt{\cal H}({\bf J})\;
    \exp\left(i\frac{}{}{\bf m}\bdot\vb{\theta} - i\,\omega_{k}t\right) \;+\;
    {\rm c.c.},
\label{eq:delta_H_Jtheta}
\end{equation}
where the parametric dependence of $\delta\wt{\cal H}$ on the Fourier indices $({\bf m},k)$ is hidden. The perturbed Vlasov distribution $\delta F$ is obtained from the perturbed Vlasov equation
\begin{equation} 
\pd{\delta F}{t} \;+\; \pd{\delta F}{\vb{\theta}}\bdot \pd{H_{0}}{\bf J} \;=\; 
\pd{\delta {\cal H}}{\vb{\theta}}\bdot\pd{F_{0}}{\bf J}, 
\end{equation} 
from which we obtain the solution for the Fourier component $\delta\wt{f}$:
\begin{equation}
    \delta\wt{F} \;=\; -\;\left(\frac{\delta\wt{\cal H}}{\omega_{k} - {\bf m} 
    \bdot\vb{\Omega}}\right)\;{\bf m}\bdot\pd{F_{0}}{\bf J},
\end{equation}
where $\vb{\Omega}({\bf J}) \equiv \partial H_{0}/\partial{\bf J}$ denotes the unperturbed orbital-frequency vector.

The quasilinear wave-particle interactions cause the Vlasov distribution 
$F_{0}({\bf J},\tau)$ to evolve on a slow time scale $\tau = \epsilon^{2}t$, represented by the quasilinear diffusion equation
\begin{eqnarray}
    \pd{F_{0}({\bf J},\tau)}{\tau} &=& \frac{1}{2}\left\langle \left\{
    \delta {\cal H},\frac{}{} \delta F \right\}\right\rangle \;=\; \frac{1}{2}
    \pd{}{\bf J}\bdot\left\langle\pd{\delta {\cal H}}{\vb{\theta}}\;\delta F
    \right\rangle \;=\; \pd{}{\bf J}\bdot\left( \sum_{{\bf m},k}
    {\bf m}\;{\rm Im}\left\langle\delta\wt{\cal H}^{*}\frac{}{}\delta\wt{F}\right\rangle \right) \nonumber \\
    &=& \pd{}{\bf J}\bdot\left[ {\rm Im}\left(\sum_{{\bf m},k}
    \frac{-\,{\bf m}{\bf m}\;|\delta\wt{\cal H}|^{2}}{\omega_{k} - {\bf m} \bdot\vb{\Omega}}\right)\bdot\pd{F_{0}}{\bf J} \right] \;\equiv\; 
    \pd{}{\bf J}\bdot\left( \vb{\sf D}_{\rm QL} \bdot\pd{F_{0}}{\bf J}\right),
    \label{eq:QL_J}
\end{eqnarray}
where $\langle\;\rangle$ includes orbital-angle averaging and wave time-scale averaging, and
the canonical quasilinear diffusion tensor
\begin{equation}
     \vb{\sf D}_{\rm QL} \;\equiv\; \sum_{{\bf m},k}{\bf m}{\bf m}\left[ \pi\frac{}{}
    \delta(\omega_{k} - {\bf m} \bdot\vb{\Omega})\right]\;|\delta\wt{H}|^{2}
    \label{eq:D_QL_J}
\end{equation}
is expressed in terms of a dyadic Fourier tensor ${\bf m}{\bf m}$, a wave-particle resonance condition obtained from the Plemelj formula
\[ {\rm Im}\left(\frac{-\,1}{\omega_{k} - {\bf m} \bdot\vb{\Omega}}\right) \;=\;
{\rm Re}\left( \frac{i}{\omega_{k} - {\bf m} \bdot\vb{\Omega}}\right) \;=\;
\pi\;\delta(\omega_{k} - {\bf m} \bdot\vb{\Omega}), \]
and the magnitude squared of the perturbed Hamiltonian Fourier component 
$\delta\wt{H}({\bf J})$, which is an explicit function of the action coordinates ${\bf J}$ and the perturbation fields (see Eq.~(63) of Ref.~\cite{Brizard_Chan:2004}, for example). We note that the perturbed Hamiltonian $\delta\wt{H}({\bf J})$ will, therefore, include terms that contain a product of an adiabatic action coordinate (such as the gyro action ${\sf J}_{\rm g}$) and a wave perturbation factor (such as $\delta B/B_{0} \sim \epsilon_{\delta}$). This means that an expansion of an adiabatic action coordinate (e.g., ${\sf J}_{\rm g} = {\sf J}_{\rm g}^{(0)} + \epsilon_{B}\,{\sf J}_{\rm g}^{(1)} + \cdots$) in the factor $|\delta\wt{H}|^{2}$ in Eq.~\eqref{eq:D_QL_J} results in a leading term of order $\epsilon_{\delta}^{2}$, followed by negligible terms of order $\epsilon_{B}\,\epsilon_{\delta}^{2} \ll \epsilon_{\delta}^{2}$. Hence, only a low-order expansion (in $\epsilon_B$) of the adiabatic action coordinates ${\bf J} \simeq 
{\bf J}_{0}$ is needed in an explicit evaluation of Eq.~\eqref{eq:D_QL_J}. In addition, we note that the form \eqref{eq:QL_J}, with Eq.~\eqref{eq:D_QL_J}, guarantees that the Vlasov entropy $S_{0} = -\,\int F_{0}\;\ln F_{0}\;d^{3}J$
\begin{equation}
\frac{dS_{0}}{dt} \;=\; -\;\epsilon^{2}\int \pd{F_{0}}{\tau}\;
(\ln F_{0} + 1)\;d^{3}J \;=\; \epsilon^{2} \sum_{{\bf m},k}\int F_{0} \left(
{\bf m}\bdot\pd{\ln F_{0}}{\bf J}\right)^{2} \pi\,
\delta(\omega_{k} - {\bf m} \bdot\vb{\Omega})\;
|\delta\wt{\cal H}|^{2}\;d^{3}J \;>\; 0
\end{equation}
satisfies the H Theorem. Lastly, we note that collisional transport in a magnetized plasma can also be described in terms of drag and diffusion in action space \cite{Bernstein:1983}.

\subsection{Local and bounce-averaged wave-particle resonances in quasilinear theory}

The canonical action-angle formalism presented in Sec.~\ref{subsec:action} unfortunately makes use of the bounce action ${\sf J}_{\rm b} = \oint p_{\|}(s)\,ds$, which is a nonlocal quantity \cite{northrop:1963}, while the drift action ${\sf J}_{\rm d} \equiv (q/2\pi c)\,\oint \psi\,d\varphi = q\psi/c$ is a local coordinate in an axisymmetric magnetic field ${\bf B} = \nabla\psi \btimes\nabla\varphi$, where the drift action is canonically conjugate to the toroidal angle $\varphi$. In our previous work \cite{Brizard_Chan:2001,Brizard_Chan:2004}, we replaced the bounce action with the guiding-center kinetic energy ${\cal E}$ in order to obtain a local quasilinear diffusion equation in three-dimensional ${\sf J}^{i} = ({\sf J}_{\rm b},{\cal E},{\sf J}_{\rm d})$ guiding-center invariant space:
\begin{equation}
\pd{F_{0}}{\tau} \;=\; \pd{}{\vb{\sf J}}\bdot\left(\vb{\sf D}_{\rm QL}\bdot
\pd{F_{0}}{\vb{\sf J}} \right) \;=\;  \frac{1}{\tau_{\rm b}} \pd{}{{\sf J}^{i}}\left( 
\tau_{\rm b}\;D_{\rm QL}^{ij}\;\pd{F_{0}}{{\sf J}^{j}} \right),   
\label{eq:QL3D}
\end{equation}
where the bounce period $\tau_{\rm b} \equiv \oint ds/v_{\|}$ is the Jacobian. In addition, the $3\times 3$ quasilinear diffusion tensor
\begin{equation}
\vb{\sf D}_{\rm QL} \;=\; \sum_{\ell,k,m} \left( \begin{array}{ccc}
\ell^{2} & \ell\,\omega_{k} & \ell\,m \\
\omega_{k}\ell & \omega_{k}^{2} & \omega_{k}m \\
m\ell & m\omega_{k} & m^{2}
\end{array}\right) \Gamma_{\ell km}
\label{eq:D_3E}
\end{equation}
is defined in terms of the Fourier indices $\ell$ (associated with the gyroangle $\zeta$) and $m$ (associated with the toroidal angle $\varphi$) and the wave frequency $\omega_{k}$, while the scalar $\Gamma_{\ell km}$ was shown in Ref.~\cite{Brizard_Chan:2004} to include the bounce-averaged wave-particle resonance condition
\begin{equation}
    \omega_{k} \;=\; \ell\,\langle\omega_{\rm c}\rangle_{\rm b} \;+\; n\,\omega_{\rm b} \;+\; m\,\langle\omega_{\rm d}\rangle_{\rm b},
    \label{eq:global}
\end{equation}
where $\langle\omega_{\rm c}\rangle_{\rm b} = (q/Mc)\,\langle B\rangle_{\rm b}$ and $\langle\omega_{\rm d}\rangle_{\rm b}$ are the bounce-averaged cyclotron and drift frequencies, respectively, and $\omega_{\rm b} = 2\pi/\tau_{\rm b}$ is the bounce frequency. Here, the bounce-average operation is defined as
\begin{equation}
\langle\;\cdots\;\rangle_{\rm b} \;\equiv\; \frac{1}{\tau_{\rm b}}\; \sum_{\sigma}\; \int_{s_{L}}^{s_{U}}\; \frac{ds}{|v_{\|}|}\;
(\cdots),
\label{eq:bounce_def}
\end{equation}
where $\sigma \equiv v_{\|}/|v_{\|}|$ denotes the sign of the parallel guiding-center velocity, and the points $s_{L,U}(\vb{\sf J})$ along a magnetic field line are the bounce (turning) points where $v_{\|}$ changes sign (for simplicity, we assume all particles are magnetically trapped). In this Section, we present a brief derivation of the quasilinear diffusion equation \eqref{eq:QL3D}, with the $3\times 3$ quasilinear diffusion tensor \eqref{eq:D_3E} and the wave-particle resonance condition \eqref{eq:global}, based on our previous work \cite{Brizard_Chan:2004}, which is presented here in the non-relativistic limit.

We begin with the linear guiding-center Vlasov equation in guiding-center phase space $(s,\varphi,\zeta;\vb{\sf J})$:
\begin{equation}
\frac{d_{0}\delta F}{dt} \;=\; \pd{\delta F}{t} \;+\; \{\delta F,\;{\cal E}\}_{\rm gc} \;=\; -\; \left\{ F_{0},\; \delta H \right\}_{\rm gc},
\label{eq:lin_Vlasov}
\end{equation}
where the perturbed Hamiltonian is a function of the guiding-center invariants $({\sf J}_{\rm b},{\cal E},{\sf J}_{\rm d})$ as well as the angle-like coordinates $(s,\varphi,\zeta)$. The unperturbed guiding-center Poisson bracket, on the other hand, is
\begin{eqnarray}
\left\{ F,\frac{}{} G \right\}_{\rm gc} &=& \pd{F}{\zeta}\;\pd{G}{{\sf J}_{\rm g}} \;-\; \pd{F}{{\sf J}_{\rm g}}\;\pd{G}{\zeta} \;+\; \pd{F}{\varphi}\;
\pd{G}{{\sf J}_{\rm d}} \;-\; \pd{F}{{\sf J}_{\rm d}}\;\pd{G}{\varphi} \;+\; \left(\frac{d_{0}F}{dt} - \pd{F}{t}\right)\pd{G}{\cal E} \;-\; \pd{F}{\cal E} \left(\frac{d_{0}G}{dt} - \pd{G}{t}\right),
\end{eqnarray}
and $d_{0}/dt = \partial/\partial t + v_{\|}\,\partial/\partial s + \omega_{\rm d}\,\partial/\partial\varphi + \omega_{\rm c}\,\partial/\partial\zeta$ denotes the unperturbed Vlasov operator ($s$ denotes the local spatial coordinate along an unperturbed magnetic-field line). Since the right side of Eq.~\eqref{eq:lin_Vlasov} is
\begin{equation}
  -\; \left\{ F_{0},\; \delta H \right\}_{\rm gc} \;=\; \pd{F_{0}}{{\sf J}_{\rm g}}\;\pd{\delta H}{\zeta} \;+\; \pd{F_{0}}{{\sf J}_{\rm d}}\;\pd{\delta H}{\varphi} \;+\;
  \pd{F_{0}}{\cal E} \left(\frac{d_{0}\delta H}{dt} - \pd{\delta H}{t}\right),
\end{equation}
we can introduce the non-adiabatic decomposition \cite{Chen_Tsai:1983}
\begin{equation}
    \delta F \;\equiv\; \delta H\;\pd{F_{0}}{\cal E} \;+\; \delta G,
\label{eq:deltaF_G}
\end{equation}
where the non-adiabatic contribution $\delta G$ satisfies the perturbed non-adiabatic Vlasov equation
\begin{equation}
    \frac{d_{0}\delta G}{dt} \;=\; \left( \pd{F_{0}}{{\sf J}_{\rm g}}\;\pd{}{\zeta} \;+\; \pd{F_{0}}{{\sf J}_{\rm d}}\;\pd{}{\varphi} \;-\; \pd{F_{0}}{\cal E}\;
    \pd{}{t}\right)\delta H \;\equiv\; \wh{\cal F}\,\delta H.
    \label{eq:nonadia_Vlasov}
\end{equation}
Next, since the background plasma is time independent and axisymmetric, and the unperturbed guiding-center Vlasov distribution is independent of the gyroangle, we perform Fourier transforms in $(\varphi,\zeta,t)$ so that Eq.~\eqref{eq:nonadia_Vlasov} becomes
\begin{equation}
    \left[ v_{\|}\,\pd{}{s} \;-\; i\,\left( \omega_{k} \;-\frac{}{} \ell\,
    \omega_{\rm c} \;-\; m\,\omega_{\rm d}\right)\right]\delta\wt{G}(s,\sigma) \;\equiv\; \wh{\cal L}\,\delta\wt{G}(s,\sigma) \;=\; i\,{\cal F}\,\delta\wt{H}(s,\sigma),
    \label{eq:nonadia_Vlasov_Fourier}
\end{equation}
where the amplitudes $(\delta\wt{G}, \delta\wt{H})$ depend on the spatial parallel coordinate $s$ and the sign $\sigma = v_{\|}/|v_{\|}| = \pm 1$, as well as the invariants $\vb{\sf J}$, while the operator $\wh{\cal F}$ becomes $i\,{\cal F}$, with
\begin{equation}
    {\cal F} \;\equiv\; \omega_{k}\;\pd{F_{0}}{\cal E} \;+\; \ell\;
    \pd{F_{0}}{{\sf J}_{\rm g}} \;+\; m\;\pd{F_{0}}{{\sf J}_{\rm d}}.
\end{equation}
In order to remove the dependence of the perturbed Hamiltonian $\delta\wt{H}$ on $\sigma$ (which appears through the combination $v_{\|}\delta\wt{A}_{\|}$), we follow our previous work \cite{Brizard_Chan:2004} and introduce the gauge $\delta\wt{A}_{\|} \equiv \partial\delta\wt{\alpha}/\partial s$ and the transformation $(\delta\wt{G},\delta\wt{H}) \rightarrow (\delta\wt{G}^{\prime},\delta\wt{K})$, where $\delta\wt{G}^{\prime} = \delta\wt{G} + i\,(q/c){\cal F}\,\delta\wt{\alpha}$ and $\delta\wt{K} = \delta\wt{H} + (q/c)\,\wh{\cal L}\,
\delta\wt{\alpha}$, so that Eq.~\eqref{eq:nonadia_Vlasov_Fourier} becomes $\wt{\cal L}\,
\delta\wt{G}^{\prime}(s,\sigma) = i\,{\cal F}\,\delta\wt{K}(s)$.

In order to obtain an integral solution for $\delta\wt{G}^{\prime}$, we now introduce the integrating factor
\begin{equation} 
\left[ v_{\|}\,\pd{}{s} \;-\; i\,\left( \omega_{k} \;-\frac{}{} \ell\,\omega_{\rm c} \;-\; m\,\omega_{\rm d}\right)\right]\delta\wt{G}^{\prime}(s,\sigma) \;\equiv\; 
e^{i\sigma\theta}v_{\|}\,\pd{}{s} \left[ e^{-i\sigma\theta}\frac{}{}\delta\wt{G}^{\prime}(s,\sigma)\right] \;=\; i\,{\cal F}\,
\delta\wt{K}(s),
\label{eq:integrating}
\end{equation}
where
\begin{equation}
    \theta(s) \;\equiv\; \int_{s_{L}}^{s} \left( \omega_{k} \;-\frac{}{} \ell\,\omega_{\rm c}(s') \;-\; m\,\omega_{\rm d}(s')\right)\; \frac{ds'}{|v_{\|}|}
    \label{eq:chi_def}
\end{equation}
is defined in terms of the lower (L) turning point $s_{L}(\vb{\sf J})$. The solution of Eq.~\eqref{eq:integrating} is, therefore, expressed as
\begin{equation}
    \delta\wt{G}^{\prime}(s,\sigma) \;=\; \delta\ov{G}^{\prime}\;e^{i\sigma\theta} \;+\; i\,\sigma\;
    e^{i\sigma\theta} \left( \int_{s_{L}}^{s} \delta\wt{K}(s')\;
    e^{-i\sigma\theta(s')} \frac{ds'}{|v_{\|}|} \right) {\cal F}
\end{equation}
where the constant amplitude $\delta\ov{G}^{\prime}$ is determined from the matching conditions $\delta\wt{G}^{\prime}(s_{L},+1) = \delta\wt{G}^{\prime}(s_{L},-1)$ and $\delta\wt{G}^{\prime}(s_{U},+1) = \delta\wt{G}^{\prime}(s_{U},-1)$ at the two turning points. At the lower turning point, the matching condition implies that $\delta\ov{G}^{\prime}$ is independent of $\sigma$. The matching condition at the upper turning point, on the other hand, is expressed as
\[ e^{i\Theta}\;\delta\ov{G}^{\prime} \;+\; \frac{i\tau_{\rm b}}{2}\;\left\langle\delta\wt{K}\frac{}{} e^{-i\theta}\right\rangle_{\rm b}\,e^{i\Theta}\;{\cal F} \;=\; e^{-i\Theta}\;
\delta\ov{G}^{\prime} \;-\; \frac{i\tau_{\rm b}}{2}\;\left\langle\delta\wt{K}\frac{}{} e^{i\theta}\right\rangle_{\rm b}\,\,e^{-i\Theta}\;{\cal F}, \]
which yields
\begin{equation} 
    \delta\ov{G}^{\prime} \;=\; -\;\frac{\tau_{\rm b}}{2} \left( \cot{\Theta}\;\left\langle \delta\wt{K}\frac{}{}\cos\theta \right\rangle_{\rm b} \;+\frac{}{} \left\langle \delta\wt{K}\frac{}{}\sin\theta 
    \right\rangle_{\rm b} \right)\;{\cal F},
    \label{eq:G_prime_sol}
\end{equation}
where
\begin{equation}
    \Theta \;\equiv\; \theta(s_{U}) \;=\; \frac{\tau_{\rm b}}{2} \left( \omega_{k} \;-\; \ell\;\langle\omega_{\rm c}\rangle_{\rm b} \;-\frac{}{} m\;\langle\omega_{\rm d}\rangle_{\rm b} \right).
    \label{eq:bigtheta}
\end{equation}
We note that $\cot\Theta$ in Eq.~\eqref{eq:G_prime_sol} has singularities at $n\pi$, which immediately leads to the resonance condition \eqref{eq:global}.

Now that the solution $\delta\wt{G}^{\prime}$ has been determined, we can proceed with the derivation of the quasilinear diffusion equation, which has been shown by Brizard and Chan \cite{Brizard_Chan:2004} to be expressed as
\begin{eqnarray}
    \pd{F_{0}}{\tau} &=& \frac{1}{\tau_{\rm b}}\pd{}{\cal E}\left[\tau_{\rm b}
    \left(\sum_{\ell,k,m}\omega_{k}\,\Gamma_{\ell km}\;{\cal F} \right) \right] +
    \frac{1}{\tau_{\rm b}}\pd{}{{\sf J}_{\rm g}}\left[\tau_{\rm b}\left(\sum_{\ell,k,m} \ell\,\Gamma_{\ell km}\;{\cal F}  \right) \right] + \frac{1}{\tau_{\rm b}}
    \pd{}{{\sf J}_{\rm d}}\left[\tau_{\rm b}\left(\sum_{\ell,k,m} m\,
    \Gamma_{\ell km}\;{\cal F}  \right) \right] \nonumber \\
     &\equiv& \frac{1}{\tau_{\rm b}} \pd{}{{\sf J}^{i}}\left( \tau_{\rm b}\;
     D_{\rm QL}^{ij}\;\pd{F_{0}}{{\sf J}^{j}} \right), 
\end{eqnarray}
which requires us to evaluate $\Gamma_{\ell km} \equiv {\cal F}^{-1}{\rm Im}\langle\delta\wt{H}^{*}\delta\wt{G}\rangle_{\rm b} = {\cal F}^{-1}{\rm Im} \langle\delta\wt{K}^{*}\delta\wt{G}^{\prime}\rangle_{\rm b}$, which is found to be expressed as
\begin{equation}
   \Gamma_{\ell km} \;=\; \frac{\tau_{\rm b}}{2}\frac{}{} {\rm Im}(-\,\cot\Theta)\; \left|\left\langle \delta\wt{K}\frac{}{} \cos\theta\right\rangle_{\rm b}\right|^{2}, 
\end{equation}
where, using the Plemelj formula with the identity $\cot z = \sum_{n=-\infty}^{\infty} (z - n\pi)^{-1}$, we finally obtain
\begin{equation}
   \Gamma_{\ell km} \;=\; \left|\left\langle \delta\wt{K}\frac{}{} \cos\theta\right\rangle_{\rm b}\right|^{2}\;\sum_{n = -\infty}^{\infty} \pi\;\delta\left( \omega_{k} \;-\frac{}{} \ell\,\langle\omega_{\rm c}\rangle_{\rm b} \;-\; n\,\omega_{\rm b} \;-\; m\,\langle\omega_{\rm d}\rangle_{\rm b}\right).
   \label{eq:Gamma_final}
\end{equation}
This expression completes the derivation of the quasilinear diffusion tensor \eqref{eq:D_3E} and the perturbed Hamiltonian $\delta\wt{K}$ is fully defined in Ref.~\cite{Brizard_Chan:2004}. We note that, in the limit of low-frequencies electromagnetic fluctuations, we also recover our previous work \cite{Brizard_Chan:2001} from Eq.~\eqref{eq:D_3E}.

We now make a few remarks concerning the bounce-averaged wave-particle resonance condition \eqref{eq:global}. First, in the case of a uniform magnetized plasma (with the drift frequency $\omega_{\rm d} \equiv 0$), we substitute the eikonal representations $\delta\wt{G} = \delta\ov{G}\,\exp(ik_{\|}s)$ and $\delta\wt{H} = \delta\ov{H}\,\exp(ik_{\|}s)$ in Eq.~\eqref{eq:nonadia_Vlasov_Fourier} and we recover the uniform quasilinear diffusion equation \eqref{eq:gcHQL_D}. Second, the bounce-averaged wave-particle resonance condition \eqref{eq:global} assumes that the waves are coherent on the bounce-time scale, which is not realistic for high-frequency (VLF), short-wavelength whistler waves \cite{Stenzel:1999,Allanson:2021}. We recover a local wave-particle resonance condition by introducing the bounce-angle coordinate $\xi(s)$ \cite{Brizard:2000}, which is defined by the equation $d\xi/ds = \omega_{\rm b}/v_{\|}$, so that $v_{\|}\,
\partial/\partial s$ in Eq.~\eqref{eq:nonadia_Vlasov_Fourier} is replaced with $\omega_{\rm b}\,\partial/\partial\xi$. Next, by introducing the bounce-angle Fourier series $\delta\wt{G} =
\sum_{n = -\infty}^{\infty} \delta\ov{G}\,\exp(in\xi)$ and $\delta\wt{H} = \sum_{n = -\infty}^{\infty} \delta\ov{H}\,\exp(in\xi)$ in Eq.~\eqref{eq:nonadia_Vlasov_Fourier}, the integral phase \eqref{eq:chi_def} is replaced by the new integral phase 
\begin{equation}
    \sigma\,\chi(s) \;=\; \sigma\,\theta(s) \;-\; n\,\xi(s) \;=\; \sigma \int_{s_{L}}^{s} \left( \omega_{k} \;-\frac{}{} \ell\,\omega_{c}(s') \;-\; m\,\omega_{\rm d}(s') \;-\; n\,
    \omega_{\rm b}\right) \frac{ds'}{|v_{\|}|}.
\end{equation}
If we now evaluate this integral by stationary-phase methods \cite{Stix:1992}, the dominant contribution comes from points $s_{0}$ along a magnetic-field line where
\begin{equation}
    0 \;=\; \chi^{\prime}(s_{0}) \;=\; |v_{\|}(s_{0})|^{-1} \left( \omega_{k} \;-\frac{}{} \ell\,\omega_{c}(s_{0}) \;-\; m\,\omega_{\rm d}(s_{0}) \;-\; n\,
    \omega_{\rm b}\right),
\end{equation}
which yields the local wave-particle resonance condition, provided $v_{\|}(s_{0}) \neq 0$ (i.e., the local resonance does not occur at a turning point).

\section{Summary}

In the present paper, we have established a direct connection between the standard reference of quasilinear theory for a uniform magnetized plasma by Kennel and Engelmann \cite{K_E} and its Hamiltonian formulation in guiding-center phase space. We have also shown that the transition to a quasilinear theory for a nonuniform magnetized plasma is greatly facilitated within a Hamiltonian formulation. The main features of a Hamiltonian formulation of quasilinear theory is that the quasilinear diffusion tensor has a simple modular dyadic form in which a matrix of Fourier indices is multiplied by a single quasilinear scalar potential, which includes the resonant wave-particle delta function. This simple modular is observed in the case of a uniform magnetized plasma, as seen in Eq.~\eqref{eq:D_HQL}, as well as in the case of a nonuniform magnetized plasma, as seen in Eq.~\eqref{eq:D_3E}. In particular, we note that the quasilinear diffusion tensor \eqref{eq:D_3E} naturally incorporates quasilinear radial diffusion as well as its synergistic connections to diffusion in two-dimensional invariant velocity space. These features are easily extended to the quasilinear diffusion of relativistic charged particles that are magnetically confined by nonuniform magnetic fields.

\appendix
\renewcommand{\theequation}{A.\arabic{equation}}

\section{\label{sec:KE_App}Kennel-Engelmann Diffusion Coefficients}

The purpose of this Appendix is to quote explicit results from the work of Kennel and Engelmann \cite{K_E}, with trivial changes in notation, and then compare them with our own representation of these results. 

\subsection{Equation (2.12) of Kennel and Engelmann}

First, Eq.~(2.12) of Kennel and Engelmann can be expressed as
\begin{equation} 
\wh{P} \;=\; \frac{q}{M} \left( \delta\wt{\bf E} + \frac{\bf v}{c}\btimes
\delta\wt{\bf B}\right)\bdot\pd{}{\bf v} \;\equiv\; e^{i(\phi - \psi)}\;\wh{P}_{+} \;+\; e^{-i(\phi - \psi)}\;\wh{P}_{-} \;+\; \wh{P}_{0},
\label{eq:2_12}
\end{equation}
where
\begin{eqnarray*}
\wh{P}_{\pm} &=& \frac{q\delta\wt{E}_{\pm}}{M}\; \frac{e^{\pm i\psi}}{\sqrt{2}} \left( \wh{G} \pm \frac{i\omega_{\|}}{v_{\bot}\,\omega}\;\pd{}{\phi} \right) \;+\;
\frac{q\,\delta\wt{E}_{\|}}{M}\;\frac{k_{\bot}}{2\omega} \left( \wh{H} \pm \frac{i\,v_{\|}}{v_{\bot}}\;\pd{}{\phi}\right), \\
\wh{P}_{0} &=& \frac{q\,\delta\wt{E}_{\|}}{M}\;\pd{}{v_{\|}} \;-\; \frac{q}{M}
\left( \delta\wt{E}_{+}\;e^{i\psi} \;-\; \delta\wt{E}_{-}\;e^{-i\psi}\right) 
\frac{i\,k_{\bot}}{2\omega}\;\pd{}{\phi},
\end{eqnarray*}
with $\omega_{\|} \equiv \omega - k_{\|}v_{\|}$, $(\delta\wt{E}_{+}, \delta\wt{E}_{-}) \equiv (\delta\wt{E}_{R}, \delta\wt{E}_{L})$, and
\begin{equation} 
\wh{G} \equiv \left(1 - \frac{k_{\|}v_{\|}}{\omega}\right)\pd{}{v_{\bot}}
\;+\; \frac{k_{\|}v_{\bot}}{\omega}\;\pd{}{v_{\|}} \;\equiv\; \pd{}{v_{\bot}} \;-\;
\frac{k_{\|}}{\omega}\;\wh{H}.
\label{eq:GH_KE}
\end{equation}
In the present work, we instead write Eq.~\eqref{eq:2_12} as
\begin{equation}
\wh{P} \;\equiv\; \Omega\left( \delta\wt{V}_{\|}\;
\pd{}{v_{\|}} + \delta\wt{V}_{\bot}\;\pd{}{v_{\bot}} + \delta\wt{\phi}\;\pd{}{\phi} \right),
\end{equation}
where the Kennel-Engelmann components are
\begin{eqnarray}
\delta\wt{V}_{\|} &=& \frac{q\,\delta\wt{\bf E}}{M\Omega}\bdot\left[ \wh{\sf z} 
\left( 1 \;-\; \frac{{\bf k}\bdot{\bf v}_{\bot}}{\omega}\right) \;+\;  \frac{k_{\|}v_{\bot}}{\omega}\left( \frac{\wh{\sf K}}{\sqrt{2}}\;e^{i(\phi - \psi)} \;+\; \frac{\wh{\sf K}^{*}}{\sqrt{2}}\;e^{-i(\phi - \psi)}\right) \right], 
\label{eq:V_par_KE} \\
\delta\wt{V}_{\bot} &=& \frac{q\,\delta\wt{\bf E}}{M\Omega}\bdot\left[ \wh{\sf z}\;
\left(\frac{k_{\bot}v_{\|}}{\omega}\right)\;\cos(\phi - \psi) \;+\; 
\left( 1 - \frac{k_{\|}v_{\|}}{\omega}\right) \left(\frac{\wh{\sf K}}{\sqrt{2}} e^{i(\phi - \psi)} \;+\; \frac{\wh{\sf K}^{*}}{\sqrt{2}} e^{-i(\phi - \psi)} 
\right) \right],
\label{eq:V_perp_KE} \\
\delta\wt{\phi} &=& \frac{q\,\delta\wt{\bf E}}{M\Omega}\bdot\left[ -\,\wh{\sf z}\;
\left( \frac{k_{\bot}v_{\|}}{\omega v_{\bot}}\right) \sin(\phi - \psi) +
\frac{i\,k_{\bot}}{\sqrt{2}\omega} \left(\wh{\sf K}^{*} - \wh{\sf K} \right) + \frac{i}{\sqrt{2}v_{\bot}} \left(1 - \frac{k_{\|}v_{\|}}{\omega} \right) \left( \wh{\sf K}\,e^{i(\phi - \psi)} -  
\wh{\sf K}^{*}\, e^{-i(\phi - \psi)} \right) \right].
\label{eq:phi_KE}
\end{eqnarray}
If we insert Faraday's Law into Eq.~\eqref{eq:delta_vpar} and use the identity \eqref{eq:id}, we obtain
\begin{eqnarray}
\delta\wt{V}_{\|} &=& \frac{q}{M\Omega} \left[ \delta\wt{\bf E} \;+\; {\bf v}_{\bot} \btimes\left(\frac{\bf k}{\omega}\btimes\delta\wt{\bf E}\right) \right]
\bdot\wh{\sf z} \nonumber \\
 &=& \frac{q}{M\Omega} \left[ \delta\wt{E}_{\|} \left( 1 \;-\; \frac{{\bf k}\bdot{\bf v}_{\bot}}{\omega}\right) \;+\; \frac{k_{\|}v_{\bot}}{\omega}\;\delta\wt{\bf E}
 \bdot\left( \frac{\wh{\sf K}}{\sqrt{2}}\;e^{i(\phi - \psi)} \;+\; \frac{\wh{\sf K}^{*}}{\sqrt{2}}\;e^{-i(\phi - \psi)}\right) \right],
\end{eqnarray}
which is identical to Eq.~\eqref{eq:V_par_KE}. If we now insert Faraday's Law into Eq.~\eqref{eq:delta_vperp} and use the identity \eqref{eq:id}, we obtain
\begin{eqnarray}
\delta\wt{V}_{\bot} &=& \frac{q}{M\Omega} \left[ \delta\wt{\bf E} \;+\;  v_{\|}\, \wh{\sf z}\btimes\left(\frac{\bf k}{\omega}\btimes\delta\wt{\bf E}\right) \right]
\bdot \frac{1}{\sqrt{2}} \left( e^{i(\phi - \psi)}\wh{\sf K} \;+\; 
e^{-i(\phi - \psi)}\wh{\sf K}^{*}\right) \nonumber \\
 &=& \frac{q\,\delta\wt{\bf E}}{M\Omega}\bdot\left[ \wh{\sf z}\;
\left(\frac{k_{\bot}v_{\|}}{\omega}\right)\;\cos(\phi - \psi) \;+\; 
\left( 1 - \frac{k_{\|}v_{\|}}{\omega}\right) \left(\frac{\wh{\sf K}}{\sqrt{2}} e^{i(\phi - \psi)} \;+\; \frac{\wh{\sf K}^{*}}{\sqrt{2}} e^{-i(\phi - \psi)} 
\right) \right],
\end{eqnarray}
which is identical to Eq.~\eqref{eq:V_perp_KE}. Lastly, if we insert Faraday's Law into Eq.~\eqref{eq:delta_phi}, and using the identity
\[ \wh{\phi} \;\equiv\; \pd{\wh{\bot}}{\phi} \;=\; \frac{i}{\sqrt{2}} \left( 
\wh{\sf K}\;e^{i(\phi - \psi)} \;-\; \wh{\sf K}^{*}\;e^{-i(\phi - \psi)}\right), \]
we obtain
\begin{eqnarray} 
\delta\wt{\phi} &=& \frac{q}{M\Omega} \left[ \delta\wt{\bf E} \;+\; {\bf v} \btimes\left(\frac{\bf k}{\omega}\btimes\delta\wt{\bf E}\right) \right]\bdot\frac{\wh{\phi}}{v_{\bot}} \;=\; \frac{q}{M\Omega} \left[ 
\delta\wt{\bf E} \left( 1 - \frac{{\bf k}\bdot{\bf v}}{\omega}\right) \;+\; \frac{{\bf k}}{\omega} \left(\delta\wt{\bf E}\bdot{\bf v}\right) \right]\bdot\frac{\wh{\phi}}{v_{\bot}} \nonumber \\
 &=& \frac{q\,\delta\wt{\bf E}}{M\Omega}\bdot\left[ -\,\wh{\sf z}\;
\left( \frac{k_{\bot}v_{\|}}{\omega v_{\bot}}\right) \sin(\phi - \psi) \;+\;
\left(1 - \frac{k_{\|}v_{\|}}{\omega}\right) 
 \frac{i}{\sqrt{2}v_{\bot}} \left( \wh{\sf K}\,e^{i(\phi - \psi)} \;-\;
 \wh{\sf K}^{*}\,e^{-i(\phi - \psi)} \right) \right. \nonumber \\
 &&\left.-\; \frac{i\,k_{\bot}}{\sqrt{2}\omega}\;\wh{\sf K}\,e^{i(\phi - \psi)} \left( \cos(\phi - \psi) \;-\frac{}{} i\;\sin(\phi - \psi)\right) \;+\; \frac{i\,k_{\bot}}{\sqrt{2}\omega}\;\wh{\sf K}^{*}\,e^{-i(\phi - \psi)} \left( \cos(\phi - \psi) \;+\frac{}{} i\;\sin(\phi - \psi)\right) \right],
\end{eqnarray}
which is identical to Eq.~\eqref{eq:phi_KE}, where we used ${\bf k}\bdot\wh{\phi} = -\,k_{\bot}\;\sin(\phi - \psi)$ and 
\[ \delta\wt{\bf E}\bdot{\bf v} \;=\; \delta\wt{E}_{\|}\;v_{\|} \;+\; \frac{v_{\bot}}{\sqrt{2}}\;\delta\wt{\bf E}\bdot\left( \wh{\sf K}\,e^{i(\phi - \psi)} \;+\; \wh{\sf K}^{*}\,e^{-i(\phi - \psi)}\right). \]

\subsection{Equation (2.19) of Kennel and Engelmann}

Equation (2.19) of Kennel and Engelmann can be expressed as
\begin{eqnarray}
\delta\wt{f} &=& \sum_{\ell,m} i\Delta_{\ell} J_{m}\;e^{i(m-\ell)(\phi - \psi)}
\left[ \frac{q\delta\wt{\bf E}}{M\Omega}\bdot\left( \frac{{\sf K}}{\sqrt{2}}\,
J_{\ell + 1} + \frac{{\sf K}^{*}}{\sqrt{2}}\,J_{\ell - 1}\right) \wh{G} +
\frac{q\delta\wt{E}_{\|}}{M\Omega}\;J_{\ell} \left(\frac{v_{\|}}{v_{\bot}}\;\wh{G}
\;+\; \frac{\Omega\,\wh{H}}{\Delta_{\ell}\,\omega v_{\bot}}\right) \right] f_{0}
\nonumber \\
 &=& \sum_{\ell,m} i\Delta_{\ell} J_{m}\;e^{i(m-\ell)(\phi - \psi)} \left(
 \delta\wt{V}_{\|\ell}\;\pd{f_{0}}{v_{\|}} \;+\; \delta\wt{V}_{\bot\ell}\;
 \pd{f_{0}}{v_{\bot}} \right),
 \label{eq:delta_f_KE}
\end{eqnarray}
where $\Delta_{\ell}$ is defined in Eq.~\eqref{eq:Delta_ell}, and the operators $\wh{G}$ and $\wh{H}$ are defined in Eq.~\eqref{eq:GH_KE}. 

First, $\delta\wt{V}_{\|\ell}$ is given by Kennel and Engelmann as
\begin{equation}
 \delta\wt{V}_{\|\ell} = \frac{q\,\delta\wt{\bf E}}{M\Omega}\bdot\left[ \wh{\sf z}\;
 J_{\ell}\; \left( 1 - \frac{\ell\Omega}{\omega}\right) \;+\; \frac{k_{\|}v_{\bot}}{\sqrt{2}\,\omega} \left( \wh{\sf K}\;J_{\ell + 1} \;+\;
 \wh{\sf K}^{*}\;J_{\ell - 1} \right) \right],
 \label{eq:V_parell_KE}
\end{equation}
If we insert Faraday's Law into Eq.~\eqref{eq:deltaV_parell}, and using the definition \eqref{eq:J_perp_l}, we obtain
\begin{eqnarray}
 \delta\wt{V}_{\|\ell} &=& \frac{q}{M\Omega} \left[ \delta\wt{E}_{\|}\;J_{\ell}
     \;-\; v_{\bot}\wh{\sf z}\btimes\left(\frac{\bf k}{\omega}\btimes\delta\wt{\bf E}
     \right)\bdot\mathbb{J}_{\bot\ell} \right] \;=\; \frac{q}{M\Omega} \left[ \delta\wt{E}_{\|} \left( J_{\ell} \;-\; \frac{v_{\bot}{\bf k}}{\omega}\bdot
     \mathbb{J}_{\bot\ell}\right) \;+\; \frac{k_{\|}v_{\bot}}{\omega}\;
     \delta\wt{\bf E}\bdot\mathbb{J}_{\bot\ell} \right] \nonumber \\
    &=& \frac{q}{M\Omega} \left[ \delta\wt{E}_{\|} \left( 1 \;-\; \frac{\ell\Omega}{\omega} \right)\;J_{\ell} \;+\; \frac{k_{\|}v_{\bot}}{\omega}\;
     \delta\wt{\bf E}\bdot\mathbb{J}_{\bot\ell} \right],
\end{eqnarray}
which is identical to Eq.~\eqref{eq:V_parell_KE}, where we used the identity 
$(v_{\bot}{\bf k}/\omega)\bdot\mathbb{J}_{\bot\ell} = (\ell\Omega/\omega)\,J_{\ell}$.

Next, $\delta\wt{V}_{\bot\ell}$ is given by Kennel and Engelmann as
\begin{equation}
\delta\wt{V}_{\bot\ell} \;=\; \frac{q\delta\wt{\bf E}}{M\Omega}\bdot\left( 
\frac{\wh{\sf K}}{\sqrt{2}}\;J_{\ell+1} + \frac{\wh{\sf K}^{*}}{\sqrt{2}}\;
J_{\ell-1}\right) \left(1 - \frac{k_{\|}v_{\|}}{\omega}\right) \;+\; \frac{q\delta\wt{E}_{\|}}{M\Omega} \frac{v_{\|}}{v_{\bot}}\; \left(\frac{\ell\Omega}{\omega}\right) J_{\ell} 
\label{eq:V_perpell_KE}
\end{equation}
If we insert Faraday's Law into Eq.~\eqref{eq:deltaV_perpell}, on the other hand, we obtain
\begin{eqnarray}
 \delta\wt{V}_{\bot\ell} &=& \frac{q}{M\Omega} \left[ \delta\wt{\bf E} \;+\;
 v_{\|}\wh{\sf z}\btimes\left(\frac{\bf k}{\omega}\btimes\delta\wt{\bf E}
     \right)\right]\bdot\mathbb{J}_{\bot\ell} \;=\; \frac{q}{M\Omega} \left[ 
     \delta\wt{E}_{\|}\;\frac{v_{\|}{\bf k}}{\omega} \;+\; \left(1 - \frac{k_{\|}v_{\|}}{\omega}\right)\;\delta\wt{\bf E}\right]
     \bdot\mathbb{J}_{\bot\ell},
\end{eqnarray} 
which is identical to Eq.~\eqref{eq:V_perpell_KE}, where we used the identity 
$(v_{\bot}{\bf k}/\omega)\bdot\mathbb{J}_{\bot\ell} = (\ell\Omega/\omega)\,J_{\ell}$.

\subsection{Equations (2.26)-(2.27) of Kennel and Engelmann}

Equation (2.26) of Kennel and Engelmann can be expressed as ${\sf D} = \sum_{\ell} \;(-i\,\Delta_{\ell})\;\wt{\bf v}_{\ell}^{*}\,\wt{\bf v}_{\ell}$, where the perturbed velocity Fourier component $\wt{\bf v}_{\ell} \equiv q\,\wt{\bf a}_{\ell}/(M\Omega)$ is given by Eq.~(2.27) as
\begin{eqnarray} 
\wt{\bf v}_{\ell} &=& \frac{q\,\delta\wt{\bf E}}{M\Omega}\bdot\left( 
\frac{\wh{\sf K}}{\sqrt{2}}\;J_{\ell+1} \;+\; \frac{\wh{\sf K}^{*}}{\sqrt{2}}\; J_{\ell+1} \right) \left[ \left(1 - \frac{k_{\|}v_{\|}}{\omega}\right)\;\wh{\bot} \;+\; \frac{k_{\|}v_{\bot}}{\omega}\;\wh{\sf z} \right] \;+\; 
\frac{q\,\delta\wt{E}_{\|}}{M\Omega}\,J_{\ell}\; \left[ \left(1 - \frac{\ell\Omega}{\omega} \right)\;\wh{\sf z} \;+\; \frac{v_{\|}}{v_{\bot}} \left(
\frac{\ell\Omega}{\omega} \right)\;\wh{\bot}\right] \nonumber \\
 &=& \frac{q}{M\Omega}\left[ \delta\wt{E}_{\|}\,J_{\ell}\;\left(1 - \frac{\ell\Omega}{\omega} \right) \;+\; \frac{k_{\|}v_{\bot}}{\omega}\;\delta\wt{\bf E}\bdot\mathbb{J}_{\bot\ell} \right]\;\wh{\sf z} \;+\; \frac{q}{M\Omega}\left[ \frac{v_{\|}}{v_{\bot}} \left(\frac{\ell\Omega}{\omega} \right)\;
 \delta\wt{E}_{\|}\;J_{\ell} \;+\; \delta\wt{\bf E}\bdot\mathbb{J}_{\bot\ell} 
 \left(1 - \frac{k_{\|}v_{\|}}{\omega}\right) \right]\wh{\bot} \nonumber \\
 &\equiv& \delta\wt{V}_{\|\ell}\;\wh{\sf z} \;+\; \delta\wt{V}_{\bot\ell}\;\wh{\bot},
\end{eqnarray}
where the Fourier components $\delta\wt{V}_{\|\ell}$ and $\delta\wt{V}_{\bot\ell}$ are given by Eqs.~\eqref{eq:V_parell_KE} and \eqref{eq:V_perpell_KE}, respectively.

\subsection{Relativistic Extension of Kennel and Engelmann}

Lerche \cite{Lerche:1968} extended the work of Kennel and Engelmann \cite{K_E} by deriving a relativistic quasilinear theory for a uniform magnetized plasma, for which the background Vlasov distribution $f_{0}({\bf p};\tau)$ is an arbitrary function of the relativistic momentum ${\bf p} = m\gamma\,{\bf v}$, where $\gamma = (1 - |{\bf v}|^{2}/c^{2}
)^{-\frac{1}{2}} = (1 + |{\bf p}/(mc)|^{2})^{\frac{1}{2}}$. In Eq.~(11) of Ref.~\cite{Lerche:1968}, the perturbed Vlasov distribution \eqref{eq:delta_f_KE} is replaced with the relativistic expression
\begin{equation}
    \delta\wt{f} \;=\; \sum_{\ell,m} i\Delta_{\ell}\; J_{m}(\lambda)\;e^{i(m-\ell)(\phi - \psi)} \left(\delta\wt{P}_{\|\ell}\;\pd{f_{0}}{p_{\|}} \;+\; \delta\wt{P}_{\bot\ell}\;
 \pd{f_{0}}{p_{\bot}} \right),
 \label{eq:delta_f_Lerche}
 \end{equation}
 where $\Delta_{\ell} = \Omega/[(k_{\|}v_{\|} - \omega) + \ell\Omega/\gamma]$, and the relativistic Fourier components $\delta\wt{P}_{\|\ell}$ and $\delta\wt{P}_{\bot\ell}$ are
 \begin{eqnarray}
 \delta\wt{P}_{\|\ell} &=& \frac{q}{\Omega} \left[ \delta\wt{E}_{\|} \left( 1 \;-\; \frac{\ell\Omega}{\gamma\,\omega} \right)\;J_{\ell}(\lambda) \;+\; \frac{k_{\|}v_{\bot}}{\omega}\;
     \delta\wt{\bf E}\bdot\mathbb{J}_{\bot\ell}(\lambda) \right], 
     \label{eq:P_par_Lerche} \\
  \delta\wt{P}_{\bot\ell} &=& \frac{q}{\Omega} \left[ \delta\wt{E}_{\|}\;\frac{v_{\|}{\bf k}}{\omega} \;+\; \left(1 - \frac{k_{\|}v_{\|}}{\omega}\right)\;\delta\wt{\bf E}\right]
     \bdot\mathbb{J}_{\bot\ell}(\lambda), 
     \label{eq:P_perp_Lerche}
 \end{eqnarray}
 where the argument of the Bessel functions in Eqs.~\eqref{eq:delta_f_Lerche}-\eqref{eq:P_perp_Lerche} is $\lambda = k_{\bot}p_{\bot}/(M\Omega) = \gamma\,k_{\bot}v_{\bot}/\Omega$, so that 
 $(v_{\bot}{\bf k}/\omega)\bdot\mathbb{J}_{\bot\ell} = (\ell\Omega/\gamma\omega)\,J_{\ell}$.
 
 \acknowledgments
 
 This work was partially funded by grants from (AJB) NSF-PHY 2206302 and (AAC) NASA NNX17AI15G and 80NSSC21K1323.

\bibliography{HQL}

\providecommand{\noopsort}[1]{#1}
\begin{thebibliography}{33}
\expandafter\ifx\csname natexlab\endcsname\relax\def\natexlab#1{#1}\fi
\expandafter\ifx\csname bibnamefont\endcsname\relax
  \def\bibnamefont#1{#1}\fi
\expandafter\ifx\csname bibfnamefont\endcsname\relax
  \def\bibfnamefont#1{#1}\fi
\expandafter\ifx\csname citenamefont\endcsname\relax
  \def\citenamefont#1{#1}\fi
\expandafter\ifx\csname url\endcsname\relax
  \def\url#1{\texttt{#1}}\fi
\expandafter\ifx\csname urlprefix\endcsname\relax\def\urlprefix{URL }\fi
\providecommand{\bibinfo}[2]{#2}
\providecommand{\eprint}[2][]{\url{#2}}

\bibitem[{\citenamefont{Kennel and Engelmann}(1966)}]{K_E}
\bibinfo{author}{\bibfnamefont{C.}~\bibnamefont{Kennel}} \bibnamefont{and}
  \bibinfo{author}{\bibfnamefont{F.}~\bibnamefont{Engelmann}},
  \bibinfo{journal}{Phys. Fluids} \textbf{\bibinfo{volume}{9}},
  \bibinfo{pages}{2377} (\bibinfo{year}{1966}).

\bibitem[{\citenamefont{Brizard and Chan}(2001)}]{Brizard_Chan:2001}
\bibinfo{author}{\bibfnamefont{A.~J.} \bibnamefont{Brizard}} \bibnamefont{and}
  \bibinfo{author}{\bibfnamefont{A.~A.} \bibnamefont{Chan}},
  \bibinfo{journal}{Phys. Plasmas} \textbf{\bibinfo{volume}{8}},
  \bibinfo{pages}{4762} (\bibinfo{year}{2001}).

\bibitem[{\citenamefont{Brizard and Chan}(2004)}]{Brizard_Chan:2004}
\bibinfo{author}{\bibfnamefont{A.~J.} \bibnamefont{Brizard}} \bibnamefont{and}
  \bibinfo{author}{\bibfnamefont{A.~A.} \bibnamefont{Chan}},
  \bibinfo{journal}{Phys. Plasmas} \textbf{\bibinfo{volume}{11}},
  \bibinfo{pages}{4220} (\bibinfo{year}{2004}).

\bibitem[{\citenamefont{Kaufman and Cohen}(2019)}]{Kaufman:2019}
\bibinfo{author}{\bibfnamefont{A.~N.} \bibnamefont{Kaufman}} \bibnamefont{and}
  \bibinfo{author}{\bibfnamefont{B.~I.} \bibnamefont{Cohen}},
  \bibinfo{journal}{J. Plasma Phys.} \textbf{\bibinfo{volume}{85}},
  \bibinfo{pages}{05850601} (\bibinfo{year}{2019}).

\bibitem[{\citenamefont{Davidson}(1972)}]{Davidson:1972}
\bibinfo{author}{\bibfnamefont{R.~C.} \bibnamefont{Davidson}},
  \emph{\bibinfo{title}{Methods in Nonlinear Plasma Theory}}
  (\bibinfo{publisher}{Academic Press}, \bibinfo{year}{1972}).

\bibitem[{\citenamefont{Stix}(1992)}]{Stix:1992}
\bibinfo{author}{\bibfnamefont{T.}~\bibnamefont{Stix}},
  \emph{\bibinfo{title}{Waves in Plasmas}} (\bibinfo{publisher}{American
  Institute of Physics}, \bibinfo{year}{1992}).

\bibitem[{\citenamefont{Sagdeev and Galeev}(1969)}]{Sagdeev_Galeev:1969}
\bibinfo{author}{\bibfnamefont{R.~Z.} \bibnamefont{Sagdeev}} \bibnamefont{and}
  \bibinfo{author}{\bibfnamefont{A.~A.} \bibnamefont{Galeev}},
  \emph{\bibinfo{title}{Nonlinear Plasma Theory}} (\bibinfo{publisher}{W. A.
  Benjamin}, \bibinfo{year}{1969}).

\bibitem[{\citenamefont{Galeev and Sagdeev}(1983)}]{Galeev_Sagdeev:1983}
\bibinfo{author}{\bibfnamefont{A.}~\bibnamefont{Galeev}} \bibnamefont{and}
  \bibinfo{author}{\bibfnamefont{R.}~\bibnamefont{Sagdeev}}, in
  \emph{\bibinfo{booktitle}{Basic plasma physics. 1}}
  (\bibinfo{publisher}{North Holland}, \bibinfo{year}{1983}).

\bibitem[{\citenamefont{Kaufman}(1972{\natexlab{a}})}]{Kaufman_JPP:1972}
\bibinfo{author}{\bibfnamefont{A.}~\bibnamefont{Kaufman}}, \bibinfo{journal}{J.
  Plasma Phys.} \textbf{\bibinfo{volume}{8}}, \bibinfo{pages}{1}
  (\bibinfo{year}{1972}{\natexlab{a}}).

\bibitem[{\citenamefont{Dewar}(1973)}]{Dewar:1973}
\bibinfo{author}{\bibfnamefont{R.}~\bibnamefont{Dewar}},
  \bibinfo{journal}{Phys. Fluids} \textbf{\bibinfo{volume}{16}},
  \bibinfo{pages}{1102} (\bibinfo{year}{1973}).

\bibitem[{\citenamefont{Dupree}(1966)}]{Dupree:1966}
\bibinfo{author}{\bibfnamefont{T.}~\bibnamefont{Dupree}},
  \bibinfo{journal}{Phys. Fluids} \textbf{\bibinfo{volume}{9}},
  \bibinfo{pages}{1773} (\bibinfo{year}{1966}).

\bibitem[{\citenamefont{Krommes}(2002)}]{Krommes:2002}
\bibinfo{author}{\bibfnamefont{J.~A.} \bibnamefont{Krommes}},
  \bibinfo{journal}{Phys. Rep.} \textbf{\bibinfo{volume}{360}},
  \bibinfo{pages}{1} (\bibinfo{year}{2002}).

\bibitem[{\citenamefont{Crews and Shumlak}(2022)}]{Crews_Shumlak:2022}
\bibinfo{author}{\bibfnamefont{D.}~\bibnamefont{Crews}} \bibnamefont{and}
  \bibinfo{author}{\bibfnamefont{U.}~\bibnamefont{Shumlak}},
  \bibinfo{journal}{Phys. Plasmas} \textbf{\bibinfo{volume}{29}},
  \bibinfo{pages}{043902} (\bibinfo{year}{2022}).

\bibitem[{\citenamefont{Dodin}(2022)}]{Dodin:2022}
\bibinfo{author}{\bibfnamefont{I.}~\bibnamefont{Dodin}}, \bibinfo{journal}{J.
  Plasma Phys.} \textbf{\bibinfo{volume}{88}}, \bibinfo{pages}{905880407}
  (\bibinfo{year}{2022}).

\bibitem[{\citenamefont{Cary and Brizard}(2009)}]{Cary_Brizard:2009}
\bibinfo{author}{\bibfnamefont{J.~R.} \bibnamefont{Cary}} \bibnamefont{and}
  \bibinfo{author}{\bibfnamefont{A.~J.} \bibnamefont{Brizard}},
  \bibinfo{journal}{Reviews of Modern Physics} \textbf{\bibinfo{volume}{81}},
  \bibinfo{pages}{693} (\bibinfo{year}{2009}).

\bibitem[{\citenamefont{Kaufman}(1972{\natexlab{b}})}]{Kaufman_PF:1972}
\bibinfo{author}{\bibfnamefont{A.}~\bibnamefont{Kaufman}},
  \bibinfo{journal}{Phys. Fluids} \textbf{\bibinfo{volume}{15}},
  \bibinfo{pages}{1063} (\bibinfo{year}{1972}{\natexlab{b}}).

\bibitem[{\citenamefont{Mahajan and Chen}(1985)}]{Mahajan_Chen:1985}
\bibinfo{author}{\bibfnamefont{S.}~\bibnamefont{Mahajan}} \bibnamefont{and}
  \bibinfo{author}{\bibfnamefont{C.~Y.} \bibnamefont{Chen}},
  \bibinfo{journal}{Phys. Fluids} \textbf{\bibinfo{volume}{28}},
  \bibinfo{pages}{3538} (\bibinfo{year}{1985}).

\bibitem[{\citenamefont{Lerche}(1968)}]{Lerche:1968}
\bibinfo{author}{\bibfnamefont{I.}~\bibnamefont{Lerche}},
  \bibinfo{journal}{Phys. Fluids} \textbf{\bibinfo{volume}{11}},
  \bibinfo{pages}{1720} (\bibinfo{year}{1968}).

\bibitem[{\citenamefont{Brizard}(1994)}]{Brizard:1994}
\bibinfo{author}{\bibfnamefont{A.}~\bibnamefont{Brizard}},
  \bibinfo{journal}{Phys. Plasmas} \textbf{\bibinfo{volume}{1}},
  \bibinfo{pages}{2460} (\bibinfo{year}{1994}).

\bibitem[{\citenamefont{Brizard}(2018)}]{Brizard:2018}
\bibinfo{author}{\bibfnamefont{A.~J.} \bibnamefont{Brizard}},
  \bibinfo{journal}{Phys. Plasmas} \textbf{\bibinfo{volume}{25}},
  \bibinfo{pages}{112112} (\bibinfo{year}{2018}).

\bibitem[{\citenamefont{Brizard and Chandre}(2020)}]{Brizard_Chandre:2020}
\bibinfo{author}{\bibfnamefont{A.~J.} \bibnamefont{Brizard}} \bibnamefont{and}
  \bibinfo{author}{\bibfnamefont{C.}~\bibnamefont{Chandre}},
  \bibinfo{journal}{Phys. Plasmas} \textbf{\bibinfo{volume}{27}},
  \bibinfo{pages}{122111} (\bibinfo{year}{2020}).

\bibitem[{\citenamefont{Northrop}(1963)}]{northrop:1963}
\bibinfo{author}{\bibfnamefont{T.~G.} \bibnamefont{Northrop}},
  \emph{\bibinfo{title}{The adiabatic motion of charged particles}}
  (\bibinfo{publisher}{Interscience Publishers}, \bibinfo{year}{1963}).

\bibitem[{\citenamefont{Lejosne and Kollmann}(2020)}]{Lejosne_Kollmann:2020}
\bibinfo{author}{\bibfnamefont{S.}~\bibnamefont{Lejosne}} \bibnamefont{and}
  \bibinfo{author}{\bibfnamefont{P.}~\bibnamefont{Kollmann}},
  \bibinfo{journal}{Space Sci. Rev.} \textbf{\bibinfo{volume}{216}}
  (\bibinfo{year}{2020}).

\bibitem[{\citenamefont{Mynick and Duvall}(1989)}]{Mynick_Duvall:1989}
\bibinfo{author}{\bibfnamefont{H.}~\bibnamefont{Mynick}} \bibnamefont{and}
  \bibinfo{author}{\bibfnamefont{R.}~\bibnamefont{Duvall}},
  \bibinfo{journal}{Phys. Fluids B: Plasma Phys.} \textbf{\bibinfo{volume}{1}},
  \bibinfo{pages}{750} (\bibinfo{year}{1989}).

\bibitem[{\citenamefont{Schulz}(1996)}]{Schulz:1996}
\bibinfo{author}{\bibfnamefont{M.}~\bibnamefont{Schulz}},
  \bibinfo{journal}{Geophysical Monograph - American Geophysical Union}
  \textbf{\bibinfo{volume}{97}}, \bibinfo{pages}{153} (\bibinfo{year}{1996}).

\bibitem[{\citenamefont{Tao et~al.}(2007)\citenamefont{Tao, Chan, and
  Brizard}}]{Tao_Chan_Brizard:2007}
\bibinfo{author}{\bibfnamefont{X.}~\bibnamefont{Tao}},
  \bibinfo{author}{\bibfnamefont{A.}~\bibnamefont{Chan}}, \bibnamefont{and}
  \bibinfo{author}{\bibfnamefont{A.}~\bibnamefont{Brizard}},
  \bibinfo{journal}{Physics of Plasmas} \textbf{\bibinfo{volume}{14}},
  \bibinfo{pages}{092107} (\bibinfo{year}{2007}).

\bibitem[{\citenamefont{Bernstein and Rowlands}(1976)}]{Bernstein:1976}
\bibinfo{author}{\bibfnamefont{I.}~\bibnamefont{Bernstein}} \bibnamefont{and}
  \bibinfo{author}{\bibfnamefont{G.}~\bibnamefont{Rowlands}},
  \bibinfo{journal}{Phys. Fluids} \textbf{\bibinfo{volume}{19}},
  \bibinfo{pages}{1546} (\bibinfo{year}{1976}).

\bibitem[{\citenamefont{Brizard and Markowski}(2022)}]{Brizard_Markowski:2022}
\bibinfo{author}{\bibfnamefont{A.}~\bibnamefont{Brizard}} \bibnamefont{and}
  \bibinfo{author}{\bibfnamefont{D.}~\bibnamefont{Markowski}},
  \bibinfo{journal}{Physics of Plasmas} \textbf{\bibinfo{volume}{29}},
  \bibinfo{pages}{022101} (\bibinfo{year}{2022}).

\bibitem[{\citenamefont{Bernstein and Molvig}(1983)}]{Bernstein:1983}
\bibinfo{author}{\bibfnamefont{I.~B.} \bibnamefont{Bernstein}}
  \bibnamefont{and} \bibinfo{author}{\bibfnamefont{K.}~\bibnamefont{Molvig}},
  \bibinfo{journal}{Phys.~Fluids} \textbf{\bibinfo{volume}{26}},
  \bibinfo{pages}{1488} (\bibinfo{year}{1983}).

\bibitem[{\citenamefont{Chen and Tsai}(1983)}]{Chen_Tsai:1983}
\bibinfo{author}{\bibfnamefont{L.}~\bibnamefont{Chen}} \bibnamefont{and}
  \bibinfo{author}{\bibfnamefont{S.-T.} \bibnamefont{Tsai}},
  \bibinfo{journal}{Plasma Phys.} \textbf{\bibinfo{volume}{25}},
  \bibinfo{pages}{349} (\bibinfo{year}{1983}).

\bibitem[{\citenamefont{Stenzel}(1999)}]{Stenzel:1999}
\bibinfo{author}{\bibfnamefont{R.~L.} \bibnamefont{Stenzel}},
  \bibinfo{journal}{J. Geophys. Res.} \textbf{\bibinfo{volume}{104}},
  \bibinfo{pages}{14379} (\bibinfo{year}{1999}).

\bibitem[{\citenamefont{Allanson et~al.}(2021)\citenamefont{Allanson, Watt,
  Allison, and Ratcliffe}}]{Allanson:2021}
\bibinfo{author}{\bibfnamefont{O.}~\bibnamefont{Allanson}},
  \bibinfo{author}{\bibfnamefont{C.~E.~J.} \bibnamefont{Watt}},
  \bibinfo{author}{\bibfnamefont{H.~J.} \bibnamefont{Allison}},
  \bibnamefont{and}
  \bibinfo{author}{\bibfnamefont{H.}~\bibnamefont{Ratcliffe}},
  \bibinfo{journal}{J. Geophys. Res.: Space Phys.}
  \textbf{\bibinfo{volume}{126}} (\bibinfo{year}{2021}).

\bibitem[{\citenamefont{Brizard}(2000)}]{Brizard:2000}
\bibinfo{author}{\bibfnamefont{A.~J.} \bibnamefont{Brizard}},
  \bibinfo{journal}{Phys. Plasmas} \textbf{\bibinfo{volume}{7}},
  \bibinfo{pages}{3238} (\bibinfo{year}{2000}).

\end{thebibliography}

\end{document}